\documentclass[epj]{svjour}

\usepackage{graphics}
\usepackage{color}

\begin{document}
 
\title{Optical conductivity of the one-dimensional
dimerized Hubbard model at quarter filling} 
\titlerunning{Optical conductivity of the one-dimensional 
dimerized Hubbard model}

\author{Holger Benthien\inst{1} \and Eric Jeckelmann\inst{2}}
 
\institute{Fachbereich Physik, Philipps-Universit\"{a}t, 
D-35032 Marburg, Germany 
\and Institut f\"{u}r Physik, 
Johannes Gutenberg-Universit\"{a}t, D-55099 Mainz, Germany
}
 
\date{Preprint: \today}
 
\abstract{
We investigate the optical conductivity in the Mott insulating phase
of the one-dimensional extended Hubbard model with alternating hopping 
terms (dimerization) at quarter band filling. Optical spectra
are calculated for the various parameter regimes using the dynamical 
density-matrix renormalization
group method. The study of limiting cases allows us to explain
the various structures found numerically in the optical conductivity
of this model. Our calculations show that the dimerization and the
nearest-neighbor repulsion determine the main features of the spectrum.
The on-site repulsion plays only a secondary role.
We discuss the consequences of our results for the theory of the
optical conductivity in the Bechgaard salts.
} 

\maketitle

\section{Introduction}
\label{sec:intro}

The electronic properties of quasi-one-dimensional charge-transfer 
salts have been intensively investigated in recent 
years~\cite{Farges,Ishiguro,Bourbonnais}.
An important example of such compounds is the family of Bechgaard
salts $({\rm TM})_2X$, where
TM is the organic molecule TMTSF
(tetramethyltetraselenafulvalene)
or TMTTF (tetramethyltetrathiafulvalene), and 
X denotes an anion such as ${\rm ClO}^-_4$, 
${\rm PF}_6^-$, ${\rm Br}^-$, etc.
These materials have highly anisotropic structures and
properties.
Therefore, it is believed that above an energy scale of a few meV
their electronic properties
can be described in first approximation
by one-dimensional models.

The one-dimensional 
Hubbard model with alternating hopping integrals
(dimerization) and a quarter-filled band has been proposed 
to describe various properties of the Bechgaard 
salts \cite{pen94,mil95,fav96,nis00,shi01,tsu01},
in particular their unusual optical 
spectrum~\cite{ped94,dre96,ves98,sch98,ves00}.
In this model the formation of a
Mott insulating ground state
is due to the interplay of the Coulomb interaction between electrons
and the lattice dimerization.
However, the relevance of this approach for
Bechgaard salts has remained controversial.
Although we know the generic features of the
low-energy optical spectrum in one-dimensional Mott
insulators~\cite{jec00,con01,gia91},
the optical conductivity of the quarter-filled
Hubbard model with dimerization has not been determined
accurately yet. Thus, no direct
comparison with the experimental
spectrum observed in Bechgaard salts has been possible.

The recent development of the dynamical density matrix 
renormalization-group (DDMRG) method~\cite{jec00,jec02} allows us
to calculate the dynamical properties of low dimensional
correlated electron systems 
with an accuracy comparable to exact diagonalizations
but for much larger system sizes.
Here, we apply the DDMRG method to the calculation of the
optical conductivity in 
the quarter-filled dimerized Hubbard model.
The model and method are briefly introduced in the next section.
Then, we present our results in Sec.~\ref{sec:results}.
Finally, we discuss the consequences of our results
for the
theory of the Bechgaard salts in Sec.~\ref{sec:discussion}.

\section{Model and Method}
\label{sec:model}

The one-dimensional extended and dimerized Hubbard model is defined
by the Hamiltonian
\begin{eqnarray}
\hat{H} &=& 
-t_1 \sum_{{\rm odd} \ l ;\sigma} \left( \hat{c}_{l,\sigma}^{\dag}
\hat{c}_{l+1,\sigma}^{\phantom{\dag}}   
+ \hat{c}_{l+1,\sigma}^{\dag}\hat{c}_{l,\sigma}^{\phantom{\dag}}
\right) \nonumber \\
&&
\label{hamiltonian}
-t_2 \sum_{{\rm even}\ l;\sigma} \left( \hat{c}_{l,\sigma}^{\dag}
\hat{c}_{l+1,\sigma}^{\phantom{\dag}} 
+ \hat{c}_{l+1,\sigma}^{\dag}\hat{c}_{l,\sigma}^{\phantom{\dag}} 
\right) \\
&&+ U \sum_{l} \hat{n}_{l,\uparrow}
\hat{n}_{l,\downarrow}  
+ V \sum_{l}(\hat{n}_l-\rho)(\hat{n}_{l+1}-\rho)  \; .  \nonumber
\end{eqnarray}
It describes fermions with spin 
$\sigma=\uparrow,\downarrow$ 
which can hop between neighboring sites representing the
highest occupied molecular orbital (HOMO) of each TM molecule. 
There are three electrons in the HOMOs of each pair
$({\rm TM})_2$, so that the band made of the HOMOs
is three-quarter filled in terms of electrons
or quarter filled in terms of holes.
We use the hole representation and keep the number of particles 
$N$ such that we have a density $\rho=N/L=1/2$ for an even number of 
lattice sites $L$.
The operator $\hat{c}^+_{l,\sigma}$ ($\hat{c}_{l,\sigma}$) creates 
(annihilates)
a hole with spin $\sigma$ at site $l$. The hole density operator
is $\hat{n}_{l,\sigma}=
\hat{c}^+_{l,\sigma}\hat{c}_{l,\sigma}$  and
$\hat{n}_l=\hat{n}_{l,\uparrow}+\hat{n}_{l,\downarrow}$ is the total 
number of holes at site $l$.
The hopping integrals $t_1 \geq t_2 \geq 0$ give rise to a 
single-particle dispersion
\begin{equation}
\epsilon(k) = \pm \sqrt{\Delta^2 \sin^2(k) +
4t^2 \cos^2(k) } 
\label{dispersion}
\end{equation}
with a total band width $4t=2t_1 + 2t_2$ and a 
(dimerization) gap $2\Delta=2(t_1-t_2)$.
The Coulomb repulsion is
mimicked by a local Hubbard interaction $U$, 
and a nearest-neighbor interaction $V$. 
The physically relevant parameter regime for Bechgaard salts is 
$U > 2V \geq 0$.
In Table~\ref{tab:1}, we show some values of the model parameters
$t_1, t_2, U$, and $V$ which have been 
proposed~\cite{mil95,fav96,nis00,duc86}
to describe various (TM)$_2$X salts.

We use open boundary conditions since density matrix renormalization
group (DMRG) algorithms
are most efficient for this type of boundary~\cite{steve,dmrgbook}.
In open chains it is important to 
use the correct form of the (non-local) Coulomb interaction
between electrons in the Hamiltonian~(\ref{hamiltonian}).
Neglecting the average density (i.e., setting $\rho=0$ in
Eq.~\ref{hamiltonian}) results in complicated edge effects
in the excitation spectrum such as the existence of
low-energy excitations localized at the chain ends. 

Two mechanisms can induce an insulating ground state in this model
at quarter filling~\cite{shi01,tsu01}.
First, the Umklapp scattering in an effectively
half-filled band [$-2t \leq \epsilon(k) \leq -\Delta$] 
due to the dimerization can lead to a 
Mott insulating state accompanied by a $4k_F$ bond order wave
(BOW), where
$k_F=\pi\rho/2=\pi/4$ is the Fermi vector.
Second, for large enough parameters $U$ and $V$
the Umklapp scattering in the quarter-filled
band [$-2t \leq \epsilon(k) \leq 2t$, 
neglecting the dimerization gap $2\Delta$]
can drive the ground state into an insulating phase
with a spontaneously broken symmetry: a $4k_F$ charge density
wave (CDW)~\cite{shi01,hir84,mil93}. 
In the family of Bechgaard salts, 
TMTSF compounds are believed to be
realizations of one-dimen\-sional Mott insulators~\cite{con01}
while TMTTF compounds are considered to be charge ordered~\cite{tsu01}
like in a CDW state.
For realistic parameters (see Table~\ref{tab:1}),
however, the system described by the
Hamiltonian~(\ref{hamiltonian}) is a Mott insulator~\cite{shi01}. 
Therefore, we will investigate the optical conductivity in 
the Mott insulating phase only.  
Since we use open boundary conditions, we observe
$2k_F$-BOW and $2k_F$- and $4k_F$-CDW
fluctuations induced by the chain ends (Friedel charge
oscillations) in the ground state.
For all the parameters $U,V,t_1,t_2$ discussed in this work,
however,
the ground state has no long-range order or broken symmetry 
but the $4k_F$-BOW induced  
by the alternating hopping terms $t_1 \neq t_2$.

\begin{table}
\caption{Model parameters (in meV) for various Bechgaard salts
from Refs.~\cite{mil95,fav96,nis00,duc86}.}
\label{tab:1}      
\begin{tabular}{llrrrr}
\hline\noalign{\smallskip}
&& $t_1$ & $t_2$ & $U$ & $V$  \\
\noalign{\smallskip}\hline\noalign{\smallskip}
(TMTSF)$_2$PF$_6$ & (Ref.~\cite{fav96}) & 250 & 225 & 1250 & 0   \\
(TMTSF)$_2$ClO$_4$ & (Ref.~\cite{nis00})& 290 & 260 & 1450 & 210 \\
(TMTTF)$_2$PF$_6$ &(Refs.~\cite{mil95,duc86}) & 135 & 95 & 945 & 380 \\
\noalign{\smallskip}\hline
\end{tabular}
\end{table}

To determine the ground state properties and to obtain
some information about excited states of the 
Hamiltonian~(\ref{hamiltonian}) we use a standard
DMRG technique~\cite{steve,dmrgbook}.
For instance, the Mott gap (also called single-particle charge gap or
charge transfer gap)
\begin{equation}
E_c = E_0(N+1)+E_0(N-1)-2E_0(N)
\end{equation}
can be calculated from the ground state energies
$E_0(N_h)$ for $N_h$ holes in the system~\cite{pen94,nis00}.
	
The linear optical absorption is proportional
to the real part $\sigma_1(\omega)$ of the optical conductivity,
which is related  
to the imaginary part of the current-current
correlation function by
\begin{equation}
\sigma_1(\omega > 0)  =  \frac{-1}{L\omega} 
{\rm Im} \ 
\langle \psi_0| \hat{J} \frac{1}{E_0+\omega + i \eta - \hat{H}} \hat{J}
|\psi_0\rangle \; .  
\label{sigma1} 
\end{equation}
Here, $|\psi_0\rangle$ is the ground state of the
Hamiltonian $\hat{H}$, $E_0$ is the ground state energy, and
$\eta \rightarrow 0^+$.
Assuming that the sites are equidistant,
the current operator $\hat{J}$ is 
\begin{eqnarray}
\hat{J} &=& 
-{\rm i}t_1 \sum_{{\rm odd} \ l ;\sigma} \left( \hat{c}_{l,\sigma}^{\dag}
\hat{c}_{l+1,\sigma}^{\phantom{\dag}}   
- \hat{c}_{l+1,\sigma}^{\dag}\hat{c}_{l,\sigma}^{\phantom{\dag}}
\right) \nonumber \\
&&
-{\rm i}t_2 \sum_{{\rm even}\ l;\sigma} \left( \hat{c}_{l,\sigma}^{\dag} 
\hat{c}_{l+1,\sigma}^{\phantom{\dag}} 
- \hat{c}_{l+1,\sigma}^{\dag}\hat{c}_{l,\sigma}^{\phantom{\dag}}
\right) \ .
\label{current}
\end{eqnarray}
With these definitions the optical conductivity
$\sigma_1(\omega)$ is given in units of $e^2a/\hbar$,
where $2a$ is the lattice constant and $e$ the charge
of a hole.
The frequency $\omega$ is given in units of $t/\hbar$.

In an open chain the optical conductivity is also related to the
imaginary part of the dipole-dipole correlation function
\begin{equation}
\sigma_1(\omega) = \frac{-\omega}{L} {\rm Im} \ 
\langle \psi_0| \hat{D} \frac{1}{E_0+\omega + i \eta - \hat{H}} \hat{D}
|\psi_0\rangle \; ,
\label{sigma2} 
\end{equation}
where the dipole operator is
\begin{equation}
\hat{D}= \sum_{l=1}^L l \left(\hat{n}_{l} - \rho \right )   
= \sum_{l=1}^L \left (l-\frac{L+1}{2} \right) \hat{n}_{l} .
\end{equation}

One can apply the DDMRG method~\cite{jec02} to chains of finite
size $L$ to compute
the optical conductivity $\sigma_1(\omega)$ with 
a finite broadening $\eta > 0$.
Thus, DDMRG yields the convolution of
$\sigma_1(\omega)$ with a Lorentzian of width $\eta$ or
the quantity defined by Eq.~\ref{sigma1} or Eq.~\ref{sigma2}
for a finite  $\eta$
(see Ref.~\cite{jec02} for more details and the advantages
of the various approaches).
The properties of the optical spectrum in the thermodynamic limit
$L \rightarrow \infty$
can be determined using a finite-size-scaling
analysis~\cite{jec02} with an appropriate broadening
\begin{equation}
\eta(L) \sim 1/L .
\label{scaling}
\end{equation}
This approach has already been successfully used to study
the optical properties of simple one-dimensional
Mott insulators (i.e, in the extended Hubbard model at half 
filling)~\cite{jec00,ess01,jec03}.
In particular, a quantitative description
has been achieved~\cite{RIXS} for the
experimental low-energy optical conductivity spectrum in 
the quasi-one-dimensional 
compound SrCuO$_2$.

Often one can use deconvolution techniques to compute a smooth
spectrum without broadening from the numerical DDMRG data
for finite $\eta$ and finite system size~\cite{geb04,uhrig}.
In this work we use a standard linear regularization method for the 
inverse problem~\cite{numrep} to deconvolve DDMRG spectra.
The deconvolution usually yields a very accurate description 
of the spectrum in the thermodynamic limit 
if it does not possess any sharp feature
(i.e., on a scale smaller than the broadening $\eta$ used
in the DDMRG calculation).
Therefore, the broadening $\eta$ used in the DDMRG calculation
sets the resolution of a
spectrum obtained through a deconvolution.  

An extension of the DDMRG method~\cite{jec02} can be used to compute
the excited states of the Hamiltonian which contribute to the
optical spectrum~(\ref{sigma1}).
We have used this method to determine the optical gap 
(i.e., the excitation energy $\omega_1$ of the lowest eigenstate 
$|\psi_1\rangle$ with a finite matrix element
$\langle\psi_1|\hat{J}|\psi_0\rangle$)
in finite chains more accurately.

Up to $m=320$
density-matrix eigenstates have been kept per block
in DDMRG calculations and up to $m=768$ 
in ground state DMRG calculations. 
Truncation errors are negligible for all results presented here. 
Thus, the accuracy of our calculations is mostly limited by the
finite broadening or resolution $\eta \sim 1/L$ imposed by 
finite system lengths.

\section{Results}
\label{sec:results}

In the absence of the non-local electron-electron interaction
($V=0$), the properties of the model~(\ref{hamiltonian}) depend
on the parameters $U$ and $\Delta$ only. (The average hopping term
$t$ just fixes the energy scale.)
First, we will investigate three limiting cases~\cite{pen94} for which
the main features in the optical conductivity $\sigma_1(\omega)$
can be easily understood:
(i) the large-dimerization limit $t_2 \ll t_1 (\Rightarrow
\Delta \approx 2t)$, $U \leq 4t_1$,
(ii) the strong-coupling limit $U \gg t_1 > t_2$,
and (iii) the weak-coupling limit $U \ll t_2 < t_1$.
Then we will discuss how the optical spectrum changes when the 
parameters
$U$ and $\Delta$ are varied between the limiting cases.
Finally, we will consider the effects
of the nearest-neighbor repulsion $V$.

\subsection{Large Dimerization}
\label{sec:dimer}

In the dimer limit $\Delta=2t$ ($t_2/t_1 \rightarrow 0$) the system
is made of independent dimers 
(i.e., pairs of nearest-neighbor sites)~\cite{pen94}.
The system eigenstates are products
of the dimer eigenstates (i.e., the eigenstates of a two-site
Hubbard model). In the ground state at quarter filling each dimer
is occupied by exactly one hole, which is localized on that dimer.
The current operator~(\ref{current}) does not couple the dimers. 
Therefore,
only intra-dimer excitations can contribute to the optical conductivity.
(They correspond to transitions from the bonding orbital 
to the anti-bonding orbital of the dimer in the limit $U=0$.)
It can be shown that $\sigma_1(\omega)$ consists of a single Dirac 
$\delta$-peak at $\omega=2t_1=4t=2\Delta$ for any $U \geq 0$.
Note, that moving one hole from a dimer to another one (inter-dimer 
excitations) yields eigenstates of the system
with an excitation energy which can be lower than $2\Delta$ and 
thus the Mott gap $E_c$ is lower
than the optical gap $\omega_1 = 2\Delta$ in that special case.
(For instance, $E_c$ vanishes as $U$ goes
to zero but $\omega_1 = 2\Delta > 0$.)

\begin{figure}
\resizebox{0.4\textwidth}{!}{\includegraphics{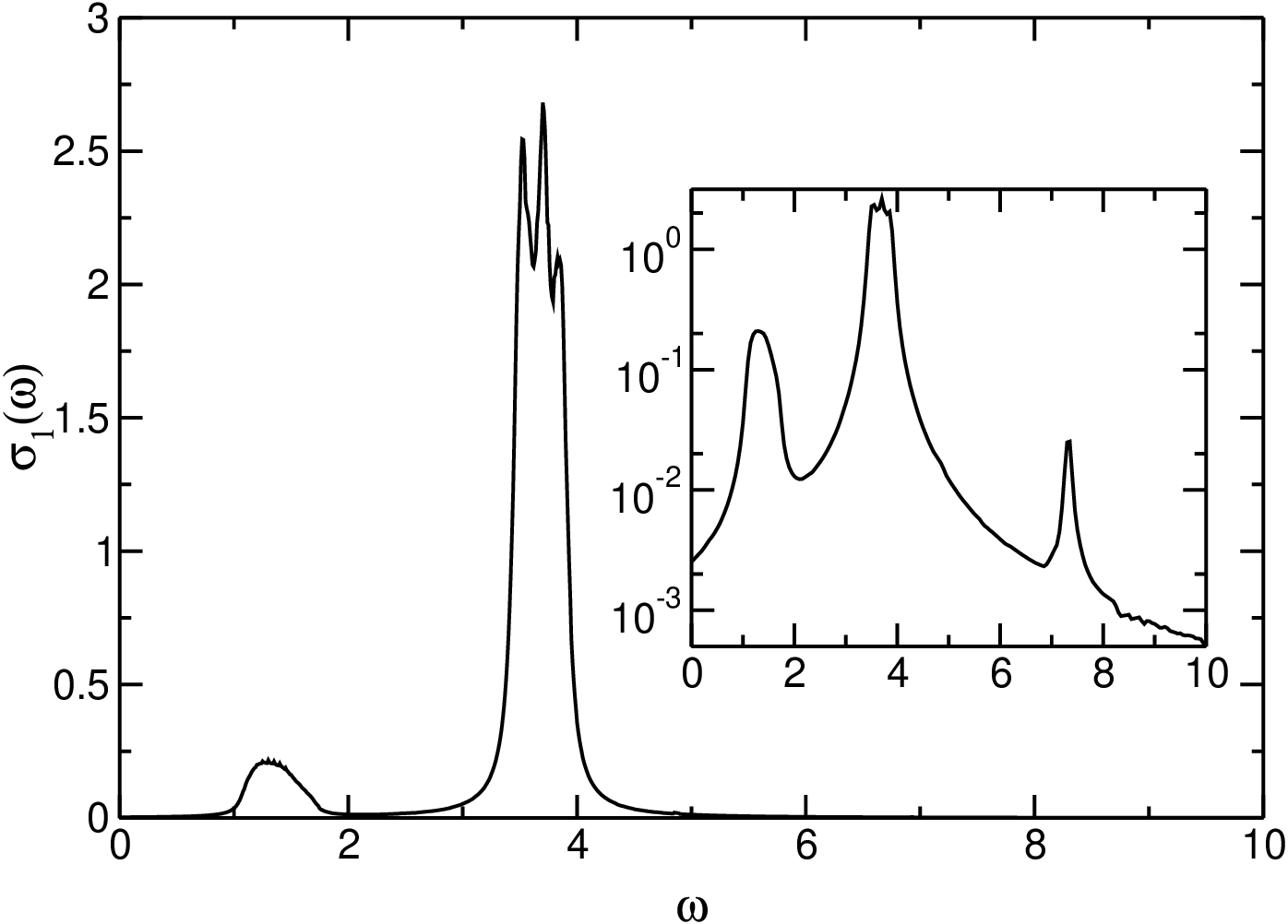}}
\caption{Optical conductivity $\sigma_1(\omega)$
in the large-dimerization limit ($\Delta=1.64t$)
for a strong effective coupling ($U=3.64t \approx 20 t_2$)
with a broadening $\eta=0.05t$ ($L=128$ sites).
Inset: same data on a logarithmic scale.
\label{fig1}       
}
\end{figure}

\begin{figure}
\resizebox{0.4\textwidth}{!}{\includegraphics{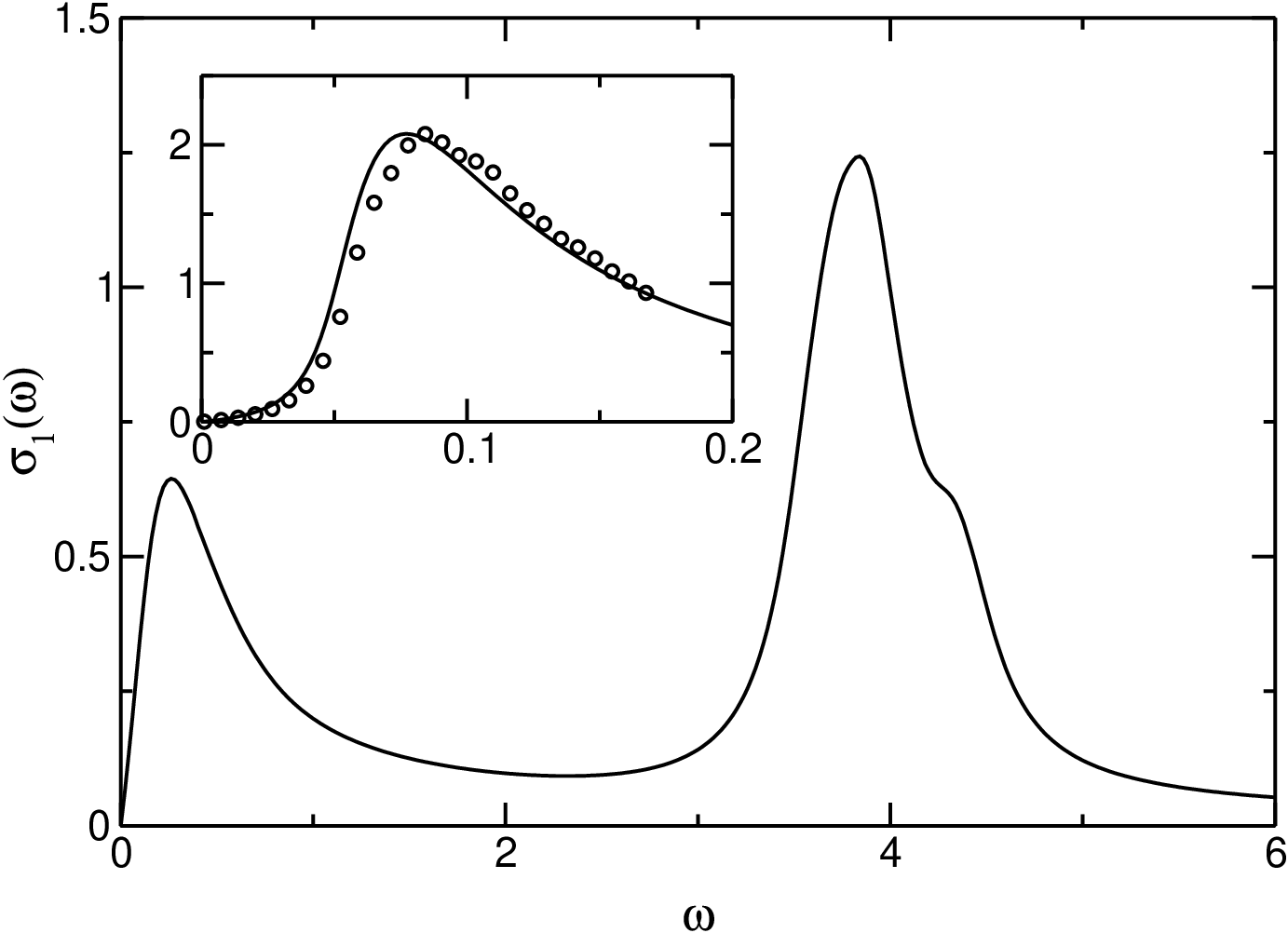}}
\caption{Optical conductivity $\sigma_1(\omega)$
in the large-dimerization limit ($\Delta=1.64t$)
for a weak effective coupling ($U=0.546t \approx 3 t_2$)
with a broadening $\eta=0.2t$ ($L=64$ sites).
Inset: high-resolution and expanded view of $\sigma_1(\omega)$ in
the low-energy region $\omega \leq 0.2t$.
DDMRG results (circles) for 
$\eta=0.0128t$ ($L=200$ sites)
and field-theoretical result~\cite{jec00} (line)
for a gap $E_c= 0.049t$ and the same broadening $\eta$.
}
\label{fig2}       
\end{figure}

We now discuss the optical excitations for a finite
inter-dimer hopping $t_2 \ll t_1$ and $U < 4t_1$. 
(For larger $U/t_1$ the spectrum is better understood starting
from the strong-coupling limit $U \gg t_1$, which is discussed in
the next section.)
For small but finite $t_2$ the dimer eigenstates become hybridized 
and build
bands of delocalized electronic states with a bandwidth $\propto t_2$.
Thus, in the optical spectrum the $\delta$-peak at $\omega=2t_1$ is 
replaced 
by an narrow absorption band (`intra-dimer' band)
of width $\propto t_2$ around $\omega = 2t_1$ 
[approximately for $2\Delta = 2(t_1-t_2) < \omega < 4t = 2(t_1+t_2)$].
Figures~\ref{fig1} and~\ref{fig2}
show the optical conductivity calculated with DDMRG
for $\Delta=1.64t \ (\Rightarrow t_2=0.18t$
and $t_1/t_2 \approx 10)$ and two different couplings
$U=3.64t$ and $U=0.546t$, respectively. 
The `intra-dimer' band contains a substantial part
of the optical weight and
is clearly visible as the strong feature 
at  $3.4 < \omega < 4.2$ in Figs.~\ref{fig1} and~\ref{fig2}.

The current operator now couples nearest neighbor dimers with a term
$\propto t_2$. Thus, for finite $t_2$ inter-dimer excitations  
also contribute to the optical spectrum.
At high energy ($\omega > 2\Delta$) these excitations
give rise to two small peaks
around $\omega = 2t_1$ and $\omega=U+2t_1$.
These features are much weaker than the `intra-dimer' band 
between $2\Delta$ and $4t$ as their optical weight
is of the order of $t_2^2/t_1$ and $t_2^2/(U+2t_1)$, respectively.
Nevertheless, the first peak is clearly visible 
on the top of the `intra-dimer' band in Fig.~\ref{fig1} and the second
peak in the inset of that figure at $\omega \approx 7.5$.  
We note that these excitations correspond to moving one particle
from the bonding orbital of a dimer to the anti-bonding
orbital of another dimer. In particular, the inter-dimer
excitation which appears in the middle of the `intra-dimer' band at
$\omega = 2t_1$ corresponds to the formation of a
triplet state on the second dimer.
Clearly this optical excitation involves both
charge and spin degrees of freedom. 

The low-energy spectrum ($\omega < 2\Delta$) is more interesting.
In the large-dimerization limit the model~(\ref{hamiltonian})
can be mapped onto a half-filled Hubbard chain with effective
parameters $t_{\rm{eff}} = t_2/2$ and
$U_{\rm{eff}} = U/2$ for $U$ small compared to $4t_1$~\cite{pen94}.
Consequently, the low-energy spectrum
is given by the optical conductivity $\sigma_1(\omega)$ 
of the  half-filled Hubbard model, 
which is known~\cite{jec00}.
For instance, for $\Delta=1.64t$ and $U=3.64t$ the effective
interaction is strong $U_{\rm{eff}}/t_{\rm{eff}} =U/t_2 \approx 20$.
Accordingly, the shape of the low-energy band ($\omega < 2$)
in Fig.~\ref{fig1}
is similar to the semi-elliptic absorption band centered
around $\omega=U_{\rm{eff}}=1.82t$ found in the strong-coupling limit
of the half-filled Hubbard model~\cite{jec00,florian2}.
Also, the optical weight in this structure is of the order of
$t^2_{\rm{eff}}/U_{\rm{eff}} = t_2^2/U$ and thus much lower
than in the `intra-dimer' band.
For $\Delta=1.64t$ and $U=0.54t$, however, 
the effective interaction is relatively weak
$U_{\rm{eff}}/t_{\rm{eff}} = U/t_2 \approx 3.3$ which corresponds
to a small Mott gap $E_c \approx 0.049t \approx 0.54 t_{\rm{eff}}$.
In that case, the optical weight in the low-energy region
$\omega < 2\Delta$ is significantly larger than for a
strong effective coupling and comparable to the
weight at higher energy, as seen in Fig.~\ref{fig2}.
Moreover, the low-energy spectrum
calculated with DDMRG (shown with a higher resolution
in the inset of Fig.~\ref{fig2}) 
agrees very well with the field-theoretical prediction
in the limit of a small Mott gap~\cite{jec00}.

\begin{figure}
\resizebox{0.4\textwidth}{!}{\includegraphics{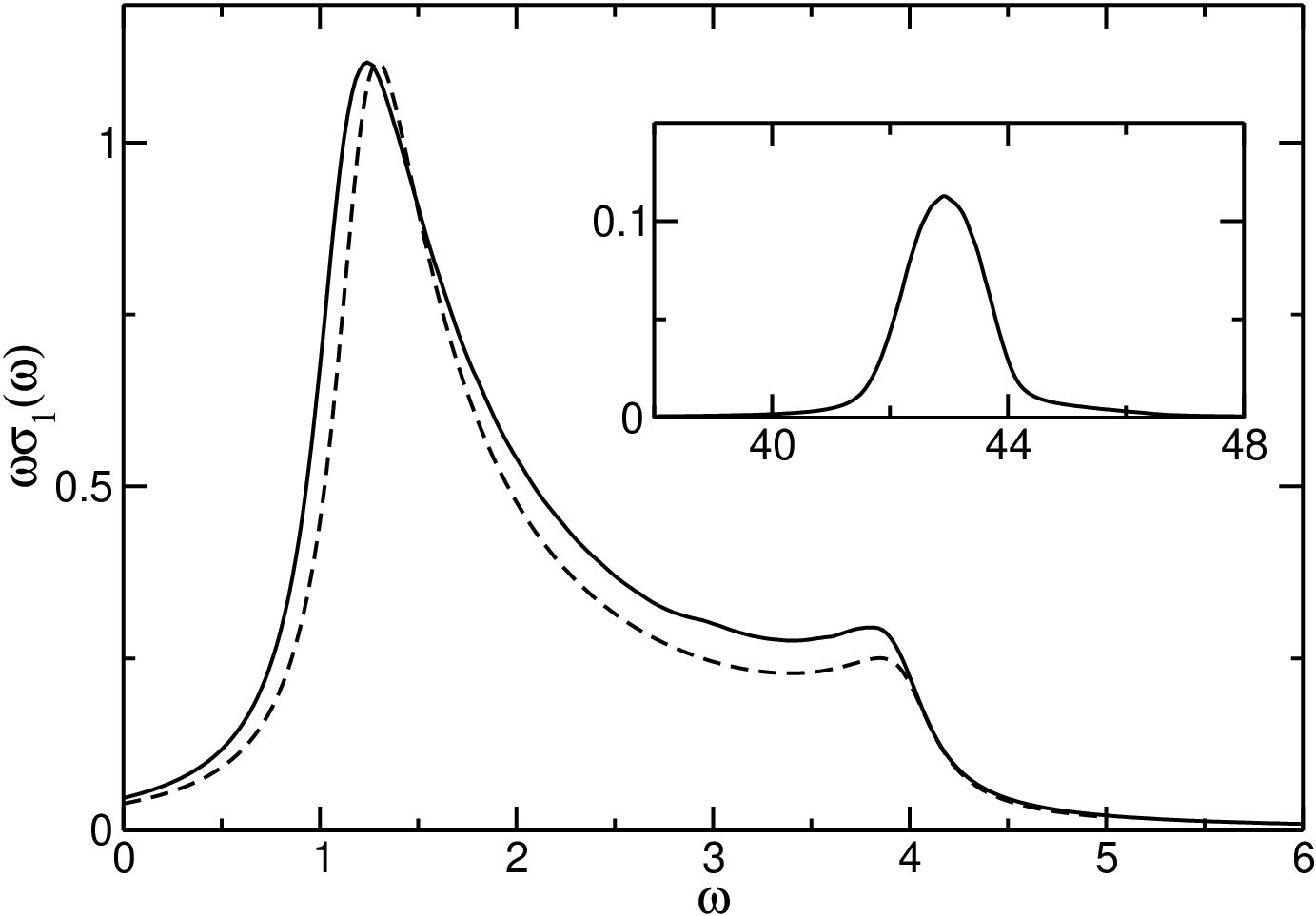}}
\caption{
Reduced optical conductivity $\omega \sigma_1(\omega)$ in the 
strong-coupling limit ($U=40t$) calculated  
with $\Delta=0.6t$ and a broadening $\eta=0.2t$ ($L=64$ sites).
The dashed line is the Peierls insulator spectrum for the
same values of $\Delta$ and $\eta$.
Inset: expanded view of the high-energy spectrum. 
}
\label{fig3}       
\end{figure}

\subsection{Strong Coupling}
\label{sec:strong}

We now turn to the strong-coupling limit $U \gg t_1 > t_2$.
In this regime the dimerization opens gaps of $2\Delta$
in the lower and upper Hubbard bands~\cite{pen94}
with band widths $4t$.
At quarter-filling
low-energy elementary charge excitations (holons in the 
lower Hubbard band) have the same dispersion~(\ref{dispersion})
as electrons in a half-filled band (Peierls) insulator.
In particular, there is a Mott gap $E_c =2\Delta$.
Neglecting the contribution of the spin degrees of freedom
to the matrix elements 
$\langle\psi_n|\hat{J}|\psi_0\rangle$, where 
$|\psi_n\rangle$ is an excited state, one expects~\cite{ped94} that the
optical conductivity $\sigma_1(\omega < U)$ is similar to that of a band
(Peierls) insulator~\cite{florian1}
\begin{equation}
\sigma_1(\omega) =  \frac{(2\Delta)^2(4t)^2}
{4 \omega^2 \sqrt{[\omega^2-(2\Delta)^2]
[(4t)^2-\omega^2]}} \; .
\label{Peierls}
\end{equation}
In Fig.~\ref{fig3} we compare this analytical result with our
numerical DDMRG data for $U=40t$ and $\Delta=0.6t$.
Both spectra have been broadened with a Lorentzian of width
$\eta =0.2t$ to facilitate the comparison.
The agreement is excellent but for a small shift $\sim t^2/U$,
which can be attributed to the finite value of $U$ used in the 
numerical 
calculations.
At high energy $\omega > U$ there is also a weak absorption band 
with a total spectral weight $\propto t^2/U$
corresponding to charge excitations from the lower to the
upper Hubbard band (see the inset of Fig.~\ref{fig3}).

\begin{figure}
\resizebox{0.4\textwidth}{!}{\includegraphics{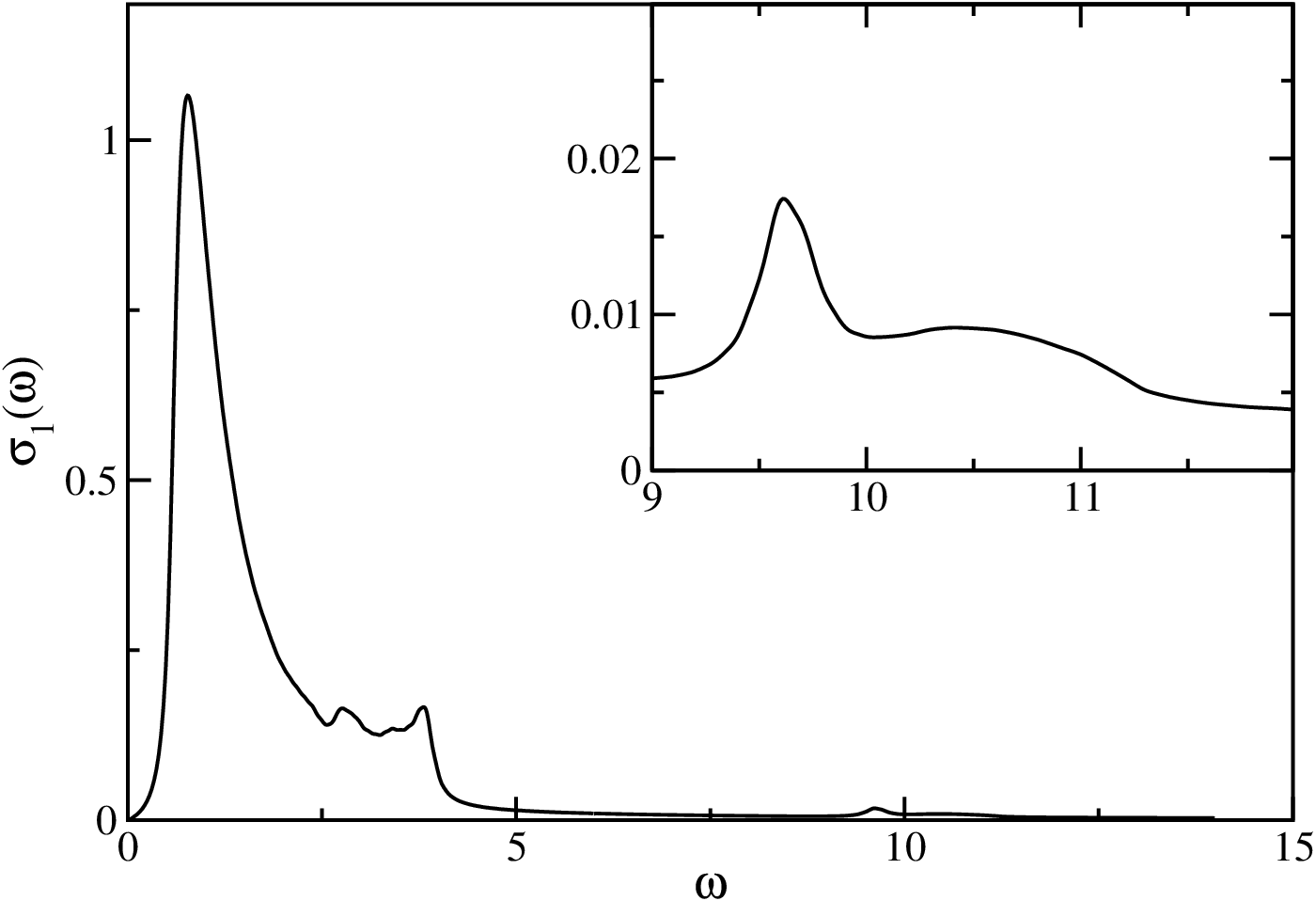}}
\caption{Optical conductivity $\sigma_1(\omega)$ for
$\Delta/t=2/3$, $U/t=20/3$, and a broadening
$\eta=0.1t$ ($L=128$ sites).       
Inset: expanded view of the high-energy spectrum.
}
\label{fig4}       
\end{figure}

The predictions of the strong-coupling theory remain
qualitatively valid for relatively weak couplings $U$.
For instance, the main features of the spectrum~(\ref{Peierls})
are clearly visible
in the optical conductivity shown in Fig.~\ref{fig4}, which
has been calculated with DDMRG for
$U=5t_1$ and $t_1/t_2=2$ (corresponding to $U/t=20/3$ and
$\Delta/t=2/3$).
There is also a weak absorption band not described by
Eq.~(\ref{Peierls}) at high energy $\omega > U$
(see inset of Fig.~\ref{fig4})
as in the strong coupling limit.
Note, however, that there are clear quantitative differences.
For instance, for these  parameters
we have found a Mott gap $E_c= 0.53t$ (in full
agreement with Ref.~\cite{pen94}), which is clearly smaller than
the strong-coupling result $2\Delta \approx 1.33t$.
          
A close inspection of the DDMRG spectra reveals
a weak structure at low frequency 
$2\Delta < \omega < 2t$ which is not explained by the simple theory
predicting the spectral form~(\ref{Peierls}).
This feature is barely visible around $\omega=3$ on the scale of
Fig.~\ref{fig3} but can be clearly seen 
as a small bump around $\omega=2.8$
in Fig.~\ref{fig4}. 
To understand this deviation from Eq.~(\ref{Peierls}) it is helpful
to analyze the spectrum as a function of the dimerization
parameter $\Delta$ for very strong coupling $U$.
For $\Delta \ll t$ (i.e., $t_1 \approx t_2$) most of the optical weight
is concentrated
in the low-energy singularity at $\omega=E_c=2\Delta$, which
in the limit $\Delta \rightarrow 0$ becomes the Drude
peak of the metallic ground state.
For $\Delta \approx 2t$ (i.e., $t_2/t_1 \ll 1$) the optical
weight becomes equally distributed between both divergences
at $\omega=E_c=2\Delta$ and $\omega=4t$.
Most of the optical weight is concentrated in this narrow band
which is the counterpart of the
`intra-dimer' band found in the large-dimerization limit
(see previous section).
Thus, the optical spectrum is dominated by a similar structure
in both the strong-coupling regime
($U \gg t_1$) with $\Delta \approx 2t$
and the large-dimerization 
limit ($\Delta \approx 2t$ but $U < 4t_1$).  
The weak spectral features, however, are quite different in both 
regimes.
In particular,
there is no optical absorption at low energy $\omega < 2\Delta$
in the strong-coupling limit ($U \gg t_1$).
The crossover from one regime to the other one 
is quite complicated and will not be discussed here because it
is not relevant for the (TM)$_2$X salts.

Nevertheless, comparing the results for large dimerization 
and those for strong coupling we observe that
the unexplained weak feature in $\sigma_1(\omega)$ for $U \gg t_1$
corresponds to the excitation involving both charge and spin
degrees of freedom at an energy $\omega=2t_1$ in the dimer limit.
Therefore, we conclude that
the spin degrees of freedom are responsible for
the (small) deviation from the simple Peierls spectrum~(\ref{Peierls})
in the large-$U$ limit.
A similarly small contribution of the spin degrees of freedom to
the optical spectrum 
has already been observed in analytical calculations~\cite{florian2}
and DDMRG simulations for the half-filled Hubbard model~\cite{jec00}.

\subsection{Weak Coupling}
\label{sec:weak}

In the weak-coupling limit $U \ll t_2 < t_1$, the low-energy
sector of the Hamiltonian~(\ref{hamiltonian})
can be mapped onto a half-filled chain with an effective (bare)
band width $2t_2$ and an effective
long-range electron-electron interaction $\propto U$
which induces a small Mott gap $E_c \ll t_2$~\cite{pen94}.
The low-energy properties $E_c \leq \omega \ll 2t_2$ of this system
should be well described by field-theoretical approaches.
In particular, one expects the optical conductivity to be
given by the field-theoretical result for 
one-dimensional Mott insulators in the small gap 
regime~\cite{jec00,con01}.
For weak coupling the optical weight must be concentrated at low energy
$\omega \sim E_c \ll t, \Delta$ as one expects that most of the spectral 
weight lies in the Drude peak when the system becomes metallic 
($E_c \rightarrow 0$).  
Therefore, the field theoretical approach describes the essential part
of the optical spectrum.

\begin{figure}
\resizebox{0.4\textwidth}{!}{\includegraphics{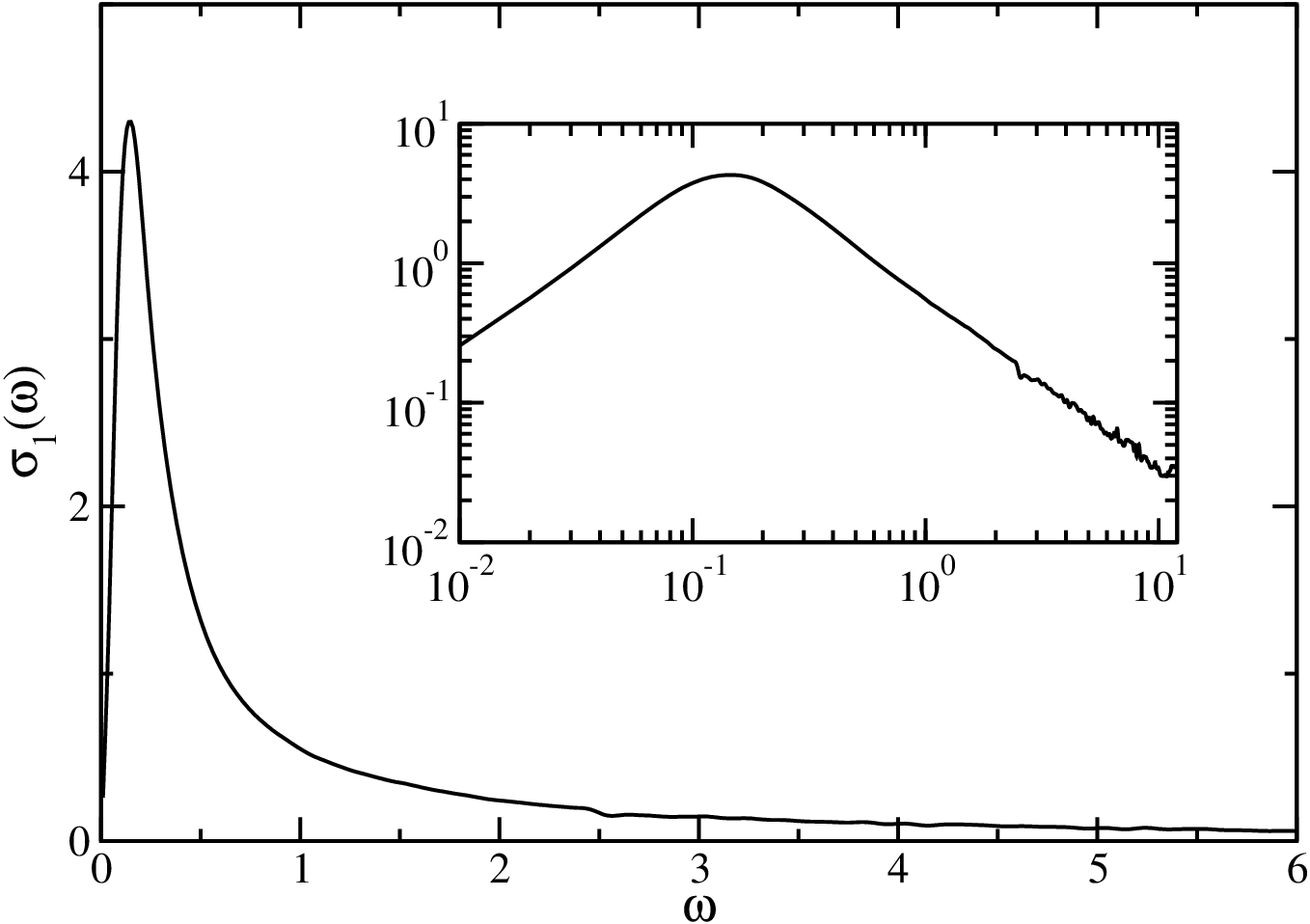}}
\caption{Optical conductivity for $\Delta=0.105t$, $U=5.263t,$
and $\eta=0.1t$ ($L=128$ sites).
Inset: same data on a double logarithmic scale.}
\label{fig5}       
\end{figure}

\begin{figure}
\resizebox{0.4\textwidth}{!}{\includegraphics{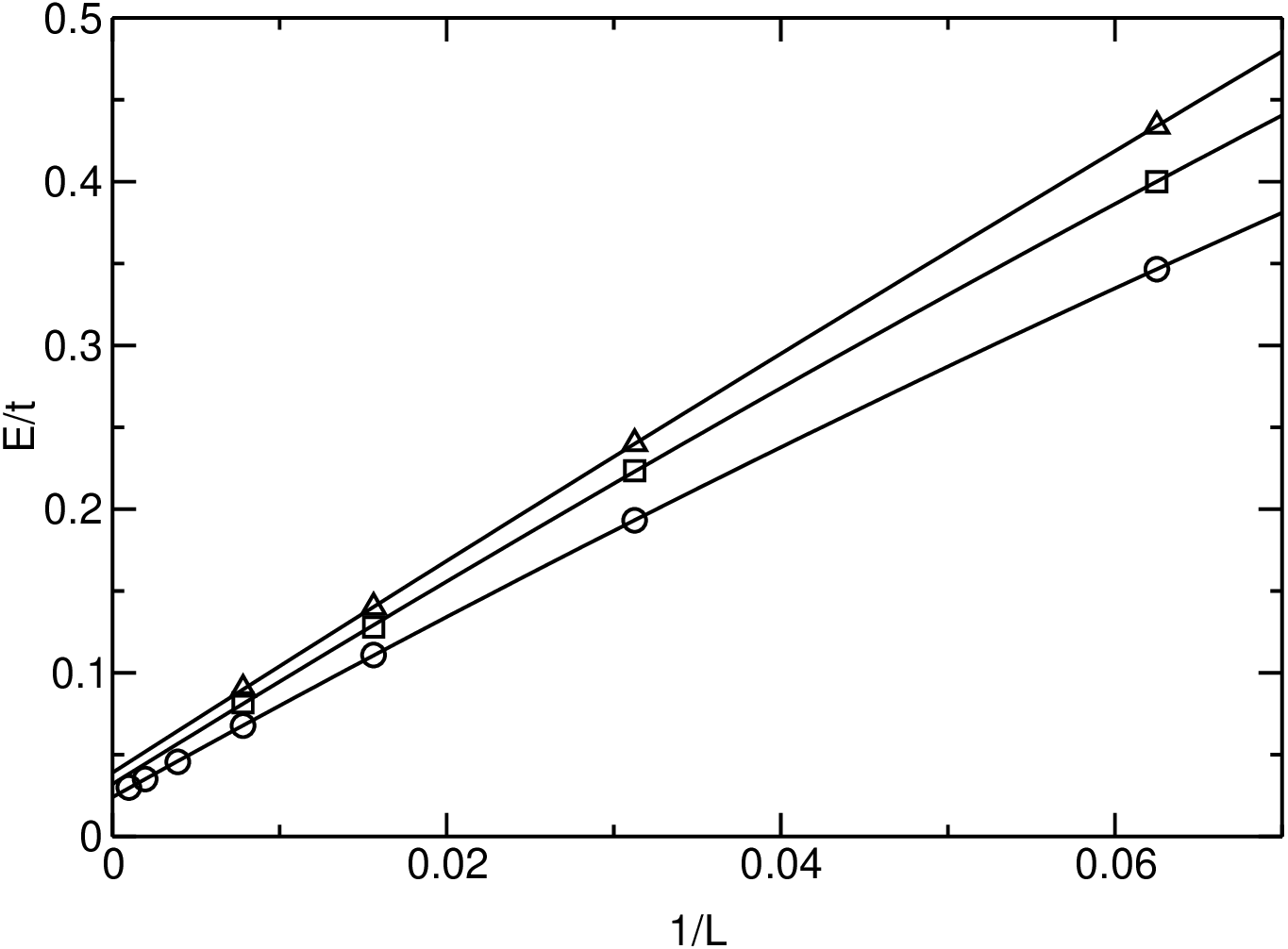}}
\caption{Mott gap $E_c$ (circle), optical gap $\omega_1$
(square), and position $\omega_{\rm max}$ of the conductivity maximum 
(triangle) as a function
of the inverse system size for $\Delta=0.105t, U=5.263t,$ and $V=0$.
Lines are quadratic fits.
}
\label{fig6}       
\end{figure}

We have performed several DDMRG calculations of $\sigma_1(\omega)$
in this weak-coupling regime.
For instance, Fig.~\ref{fig5} shows the optical conductivity
calculated for $U=5.263t$ and $\Delta=0.105t$
(corresponding to $t_2/t_1 =0.9$).
Clearly, the optical weight is concentrated in a sharp peak at low
frequency as expected (the long tail at high frequency is
mostly due to the broadening $\eta=0.1t$).
For these parameters the Mott gap is $E_c \approx 0.03t$ in the
thermodynamic limit and the optical gap converges to the same value
as seen in Fig.~\ref{fig6}.
The broadening of the DDMRG spectrum 
also results in an apparent shift of the peak position
[i.e., the maximum $\omega_{\rm max}$ of $\sigma_1(\omega)$] 
to higher frequencies.
In the limit of an infinite chain and with the scaling~(\ref{scaling})
one finds (see Fig.~\ref{fig6}) that
$\omega_{\rm max}$ approaches a value ($0.04t$) only
slightly larger
than the Mott gap $E_c$ in agreement with the field theory
prediction~\cite{jec00,con01}.
As most of the spectral weight is concentrated on a scale
$\omega \sim E_c$ comparable or smaller than our typical resolution
$\eta$, it is not possible to make a quantitative comparison
between field theory and numerical results for the spectral  lineshape.
Nevertheless, our DDMRG results are always qualitatively compatible
with field-theoretical predictions for the behavior of
$\sigma_1(\omega)$ at frequencies of the order of the charge 
gap $E_c$.

A field-theoretical analysis~\cite{gia91} of the optical conductivity
in one-dimensional Mott insulators predicts that
the leading asymptotic behavior at high frequency
$\omega \gg E_c$ is 
\begin{equation}
\sigma_1(\omega) \sim \omega^{-\alpha} ,
\label{asymptotic}
\end{equation}
where the exponent $\alpha \geq 1$ depends on the interaction
strength.
This power-law behavior
is a good approximation at extremely high frequencies 
$\omega \sim 10^2 E_c$ only~\cite{con01} and the field theory
approach is only valid for energies much smaller than the 
effective band width $\sim 2t$.
Thus, in the lattice model~(\ref{hamiltonian}) one can observe
such an asymptotic behavior in the limit of small Mott gaps
only (i.e., for $E_c \ll \omega \ll t$).
Moreover, the high-frequency behavior can be modified by 
various processes which are neglected in the field-theoretical approach
but generate additional optical transitions for $\omega > E_c$,
such as inter-band transitions at $\omega \geq 2\Delta$ or
transitions between the lower and upper Hubbard band around
$\omega = U$.
For instance, we have seen in Sec.~\ref{sec:dimer}
that in the (effective) weak-coupling regime
of the dimerized limit  the low-frequency
spectrum $\omega \sim E_c \ll t$
can be described with field theory but the high-frequency
spectrum is dominated by the inter-dimer
excitations around $\omega \approx 2\Delta$ (see Fig.~\ref{fig2}).
Obviously, a power-law behavior cannot be observed
for $E_c \ll \omega \ll t$ in that case.
In practice, only the weak-coupling regime of the 
Hamiltonian~(\ref{hamiltonian}) seems to fulfill both conditions 
necessary
for the occurrence of the asymptotic power-law behavior of 
$\sigma_1(\omega)$: (i) a small Mott gap $E_c \ll t$ and (ii)
no other optical excitation in the relevant
frequency range.   

We have found that DDMRG spectra for finite broadening $\eta$
and system size often decay as a power-law at high frequency.
In the inset of Fig.~\ref{fig5} one clearly sees such a behavior
with an exponent $\approx -1.2$ 
for $0.2t < \omega < 10t$ corresponding to $7E_c < \omega < 330 E_c$.
(In Sec.~\ref{sec:coupling} we will see that the high-frequency 
spectrum is 
actually dominated by another feature explained by a  
strong-coupling analysis.) 
However, the exponent and the range over which the 
power-law behavior
can be observed depend on the method used to broaden
the spectrum in the DDMRG calculation (see Sec.~\ref{sec:model}),
the broadening $\eta$, and even the system size. Therefore,
this power-law decay is probably an artifact of our numerical
approach.
This effect can easily be understood if one assumes that most of the
optical weight is concentrated in a sharp structure at $\omega \sim 
E_c$.
The broadening of this structure creates a broad tail which 
decreases asymptotically as $\eta A_1 \omega^{-\beta}$,
where the exponent $\beta$ lies between 1 and 3
(depending on the precise broadening technique
used in the DDMRG simulation) and the coefficient $A_1$
is proportional to the total optical weight.
Obviously, this artificial broad tail hides
an asymptotic behavior 
$\sigma_1(\omega)\sim A_2 \omega^{-\alpha}$
for $\beta < \alpha$.
It can also hide the asymptotic behavior
of $\sigma_1(\omega)$
up to relatively large frequencies 
for $\beta > \alpha$
if the high-frequency spectrum contains
only a small fraction of the total optical weight
(i.e., $A_2 \ll \eta A_1$).
We think that this effect is responsible for the power-law observed in 
our DDMRG spectra with a finite broadening $\eta$.

To determine the true asymptotic behavior of $\sigma_1(\omega)$
we have tried to deconvolve the DDMRG spectra for finite $\eta$ in order
to obtain spectra for $\eta=0$ in the thermodynamic 
limit~\cite{geb04}.
The resulting spectra do {\it not} show a power-law behavior in 
any significant range of frequencies.
Unfortunately, the accuracy of the deconvolved spectra is very poor 
at high frequencies because
our deconvolution technique (linear regularization
approach for an inverse problem~\cite{numrep})
does not work well when 
the spectrum is dominated by sharp structures as in Fig.~\ref{fig5}.

In summary, we have not been able to determine the asymptotic
behavior of $\sigma_1(\omega)$ in the weak-coupling regime. 
While some of the raw DDMRG data clearly
exhibit a power-law behavior at high 
frequency, we think that this is an artifact of our method. 
We can not confirm (or refute) the validity of the field-theory
prediction~(\ref{asymptotic}) for the lattice model~(\ref{hamiltonian})
investigated here.
Nevertheless, our investigation leads us to conclude that
an asymptotic power-law behavior~(\ref{asymptotic}) can
occur only in the weak-coupling regime. 
Moreover, the optical weight at high-frequency
(i.e., in the asymptotic tail) can only be a small fraction
of the total optical weight, which is concentrated just above
the gap $E_c$.

\begin{figure}
\resizebox{0.4\textwidth}{!}{\includegraphics{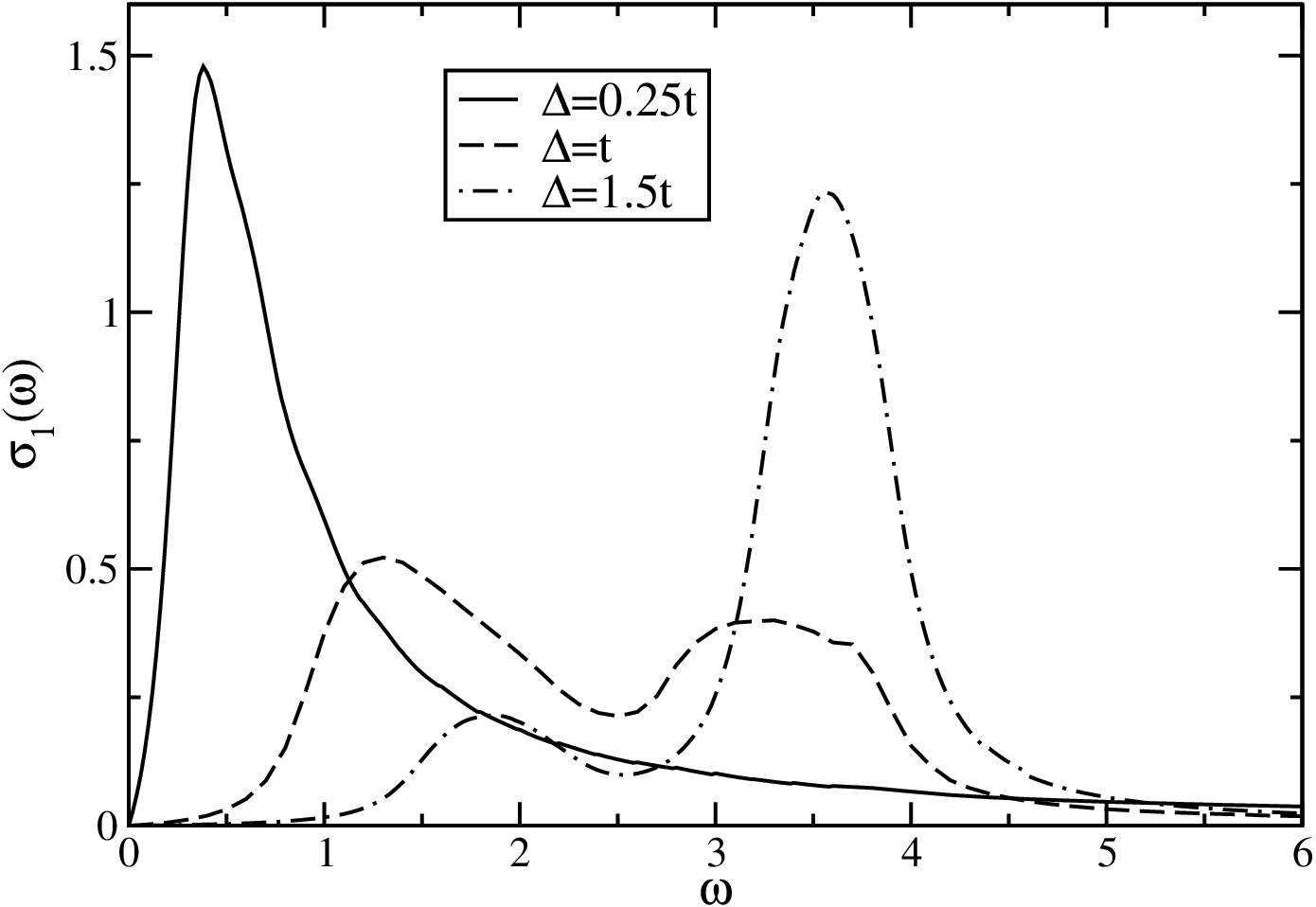}}
\caption{Optical conductivity $\sigma_1(\omega)$ for $U=6t$
and $\eta = 0.2t$ ($L=32$) for various
dimerizations $\Delta$.  
}
\label{fig7}       
\end{figure}

\subsection{From Small to Large Dimerization}
\label{sec:dimerization}

Although the nature of the optical excitations greatly differs
in the strong and weak coupling limits, we have found
that the evolution of the optical
spectrum with $\Delta$ is qualitatively similar for all values of
$U > 0$. 
In the strong-coupling limit the spectrum is given
by Eq.~(\ref{Peierls}) and thus the evolution
of $\sigma_1(\omega)$ with $\Delta$ can be described accurately.
When the dimerization is weak ($\Delta \ll t$), 
most of the optical
weight is concentrated just above the first singularity
at the spectrum onset  
$\omega = E_c = 2\Delta$ and the second singularity at $\omega=4t$
carries very little weight.
As $\Delta$ increases, the optical weight is progressively
transfered from the low-energy singularity
to the high-energy one until the spectral weight becomes
equally distributed between both singularities as 
one reaches the large dimerization limit
$\Delta \rightarrow 2t$.
Simultaneously, the first singularity moves to higher energy
as $\Delta$ increases
and ultimately merges with the second (fixed) one as
$\Delta$ reaches  $2t$.
Therefore, we observe both a transfer of optical weight from a
low-energy structure around $\omega = E_c$ to a high-energy 
structure around $\omega=4t$
and a shift of the low-energy structure toward higher
excitation energies as $\Delta$ increases.

Away from the strong-coupling limit Eq.~(\ref{Peierls})
is not an accurate description of the optical spectrum. 
Nevertheless,
our DDMRG calculations show a qualitatively similar evolution 
of the spectrum as a function of $\Delta$ 
for all values of $U$ that we have analyzed (i.e., down to $U=t$).
For instance, Fig.~\ref{fig7} shows the optical conductivity
calculated with DDMRG for $U=6t$ and several values of $\Delta$.
We clearly see both the optical weight transfer from the low-energy
peak to the high-energy structure and the shift of the low-energy
peak toward higher energy as $\Delta$ increases.
We note, however, that 
the low-energy peak is close to the Mott gap
$E_c$ for small $\Delta$ only. 
For larger $\Delta$ the position of the first peak
moves away from $E_c$ (at least when $U$ is not too large)
contrary to the strong-coupling result~(\ref{Peierls}). 
As a result, the dominant features in $\sigma_1(\omega)$ can
lie well above the Mott gap $E_c$ as already shown
for the large-dimerization limit in Sec.~\ref{sec:dimer}.
The high-energy structure always lies at an energy close to $4t$
for all $U$ and $\Delta$ but its weight can become so small that
it is no longer visible such as in the weak-coupling limit
(see Fig.~\ref{fig5}).

\begin{figure}
\resizebox{0.4\textwidth}{!}{\includegraphics{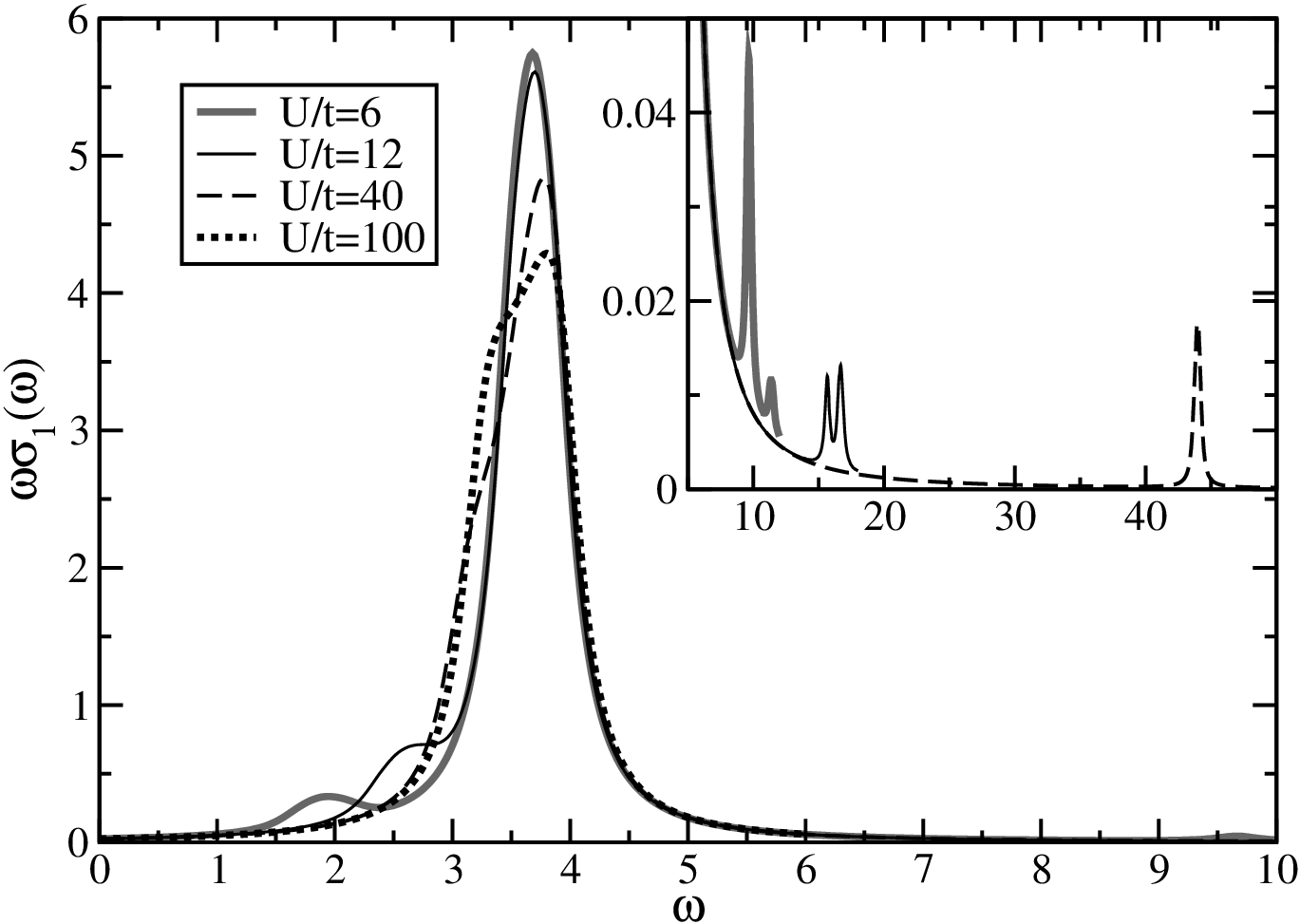}}
\caption{Reduced optical conductivity $\omega\sigma_1(\omega)$
calculated with  $\eta=0.2t$ ($L=64$) for $\Delta=1.64t$ and
various values of $U$. Inset:
expanded view of the high-frequency conductivity spectrum.
}
\label{fig8}       
\end{figure}

\subsection{From Weak to Strong Coupling}
\label{sec:coupling}

For a given dimerization $\Delta < 2t$ the strength $U$ of the local
Coulomb interaction significantly modifies the Mott gap $E_c$,
which is equal to the optical gap for $\Delta < 2t$ and $V=0$.
However, it has little effect on the main features of the optical 
spectrum and just modifies the fine structure.
For instance, Fig.~\ref{fig8} shows the reduced optical conductivity 
calculated with DDMRG for $\Delta=1.64t$ and various values of $U$.
For $U \leq 4t$ the optical spectrum is well explained by the analysis 
of the large-dimerization limit (Sec.~\ref{sec:dimer}). 
It consists of a 
strong structure at $\omega \approx 2t_1 = 3.64t$ (the intra-dimer band)
and weaker features at $\omega \approx E_c$ and $\omega \approx 2t_1+U$
(see Fig.~\ref{fig1}).
When $U$ increases, the gap becomes larger and correspondingly we 
observe a progressive shift of the low-frequency weak structure toward 
the strong peak in Fig.~\ref{fig8}.
Simultaneously, the high-frequency weak feature moves to higher energies
in agreement with the relation $\omega \approx 2t_1+U$ (see the inset
of Fig.~\ref{fig8}).
However, the strong dominant structure remains largely unaffected by 
the variation of the coupling $U$.

Nevertheless, the analysis of the optical conductivity for changing $U$
yields an interesting result. 
As discussed in Sec.~\ref{sec:weak} in the weak-coupling approach
the local interaction term
is responsible for an (effective) interaction $\propto U$ which splits 
the (effectively)
half-filled band [defined by the lower part of the single-particle
dispersion~(\ref{dispersion})] in two (effective) Hubbard bands 
separated by a gap $E_c$.
Excitations from the lower to the upper effective Hubbard bands 
significantly contribute to
the optical spectrum on the energy scale $\omega=E_c$.
In the strong-coupling approach the local interaction term splits the 
full
band defined by the dispersion~(\ref{dispersion}) in two (full)
Hubbard bands separated
by a gap $\sim U$. In that case, excitations from the lower to the 
upper full
Hubbard bands contribute to the optical spectrum around $\omega=U$.
Our calculations show that both features can be seen
(for instance, in Figs.~\ref{fig1} and~\ref{fig8}) for a given value 
of $U$. 
Therefore, we conclude that, at least for some parameters ($\Delta, U$)
of the model~(\ref{hamiltonian}),
the low-frequency part of the optical spectrum is explained by
weak-coupling (i.e., field-theoretical) approaches while the 
high-frequency part is explained by a strong-coupling 
analysis.

\begin{figure}
\resizebox{0.4\textwidth}{!}{\includegraphics{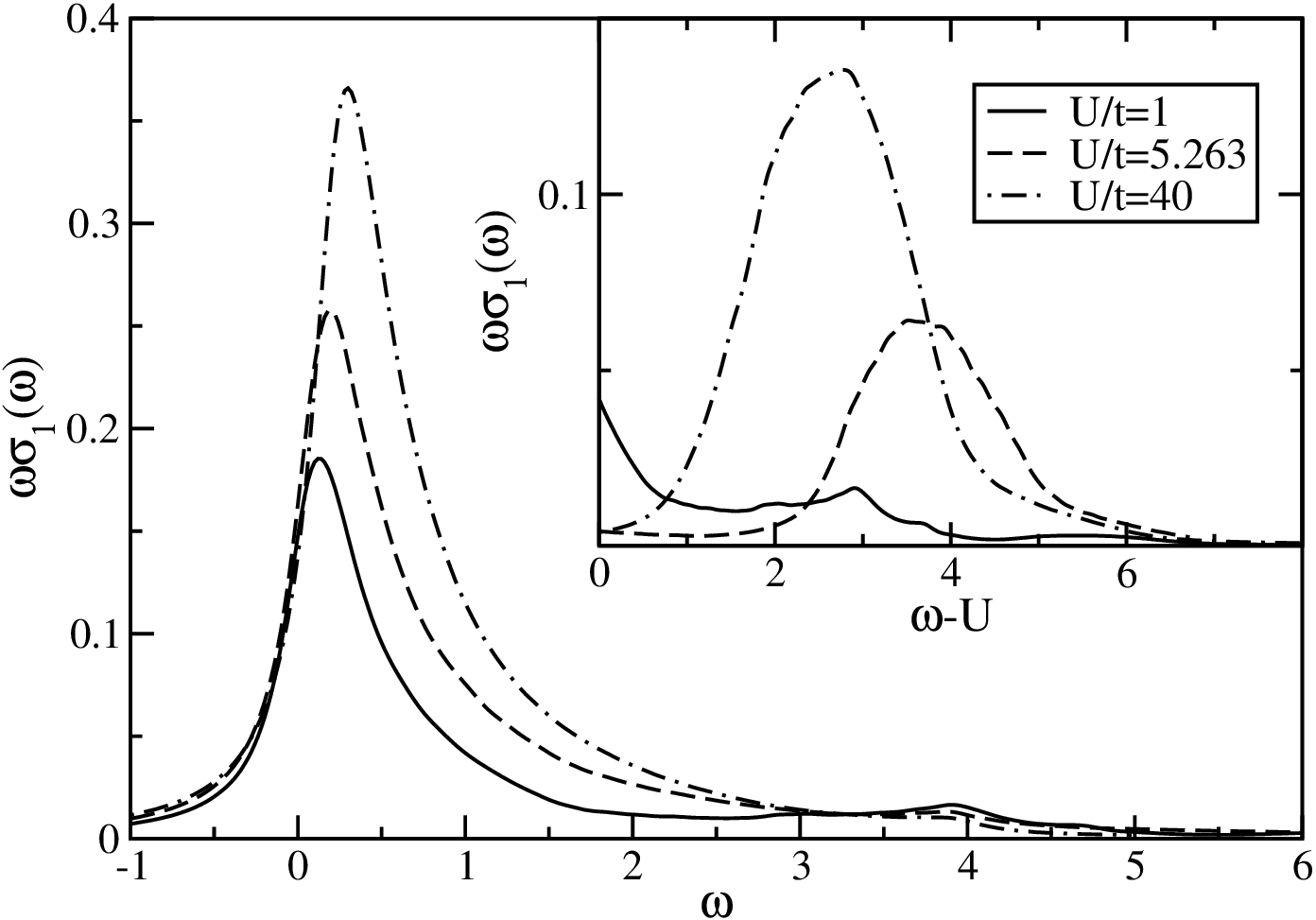}}
\caption{Reduced optical conductivity $\omega\sigma_1(\omega)$
calculated with  $\eta=0.2t$ ($L=64$) for $\Delta=0.105t$ and
various values of $U$. Inset:
expanded view of the high-frequency conductivity spectrum
as a function of $\omega-U$.
}
\label{fig9}       
\end{figure}

As a second example of the optical spectrum evolution with $U$,
we show the reduced optical conductivity  calculated
with DDMRG for a small dimerization $\Delta=0.105t$ and various 
interaction strengths $U$ in Fig.~\ref{fig9}.
For $U=t$ the system is in the weak-coupling limit (discussed in 
Sec.~\ref{sec:weak})
and most of the spectral weight is concentrated in a peak close to 
$\omega = E_c$.
A second weaker feature is visible at about $\omega =4t$ (see the inset
of Fig.~\ref{fig9}) and corresponds to transitions from the bottom
to the top of the single-particle band~(\ref{dispersion}).
For larger $U$ the optical weight remains concentrated in the low-energy
peak at $\omega \approx E_c$. This peak moves to slightly higher
energies
and appears to broaden because the energy scale set by the gap
$E_c$ increases with $U$ until it reaches $2\Delta$ for
$U \rightarrow \infty$ as discussed in Sec.~\ref{sec:strong}
but its shape is not significantly changed by the variation of $U$.
The feature at $\omega \approx 4t$ disappears for $U > 4t$ but 
another  weak feature  becomes visible around $\omega \approx U$ 
for strong enough coupling $U$ (see the inset of Fig.~\ref{fig9}).
Again this corresponds to optical excitations from the lower to the
upper (full) Hubbard bands.

We note that for $U=5.263t$ this contribution to the optical
spectrum is already clearly visible around $\omega-U = 4t$ in
the inset of Fig.~\ref{fig9}.
This result contradicts the apparent asymptotic power-law 
decrease~(\ref{asymptotic})
discussed previously for the same parameters (see Fig.~\ref{fig5}).
This discrepancy is due to the different broadening methods 
and data representation used for Figs.~\ref{fig5} and~\ref{fig9}.
It confirms that the asymptotic power law found
in some of our spectra for weak couplings are probably an artifact
of the broadening used in the DDMRG calculation.  
This also illustrates how difficult it is to observe
the field-theoretical 
predictions~(\ref{asymptotic}) for the asymptotic behavior of
$\sigma_1(\omega)$ in the 
lattice model~(\ref{hamiltonian}) because optical transitions
neglected in the field theory approach significantly
contributes to the high-frequency spectrum.

In summary,
our calculations for the model~(\ref{hamiltonian}) with $V=0$
show that the distribution of the optical
weight is essentially determined by the dimerization amplitude
$\Delta$.
For a fixed $\Delta$ 
only the fine structure of the optical spectrum and the energy
scale set by the gap $E_c$ depend significantly on $U$.

\begin{figure}
\resizebox{0.4\textwidth}{!}{\includegraphics{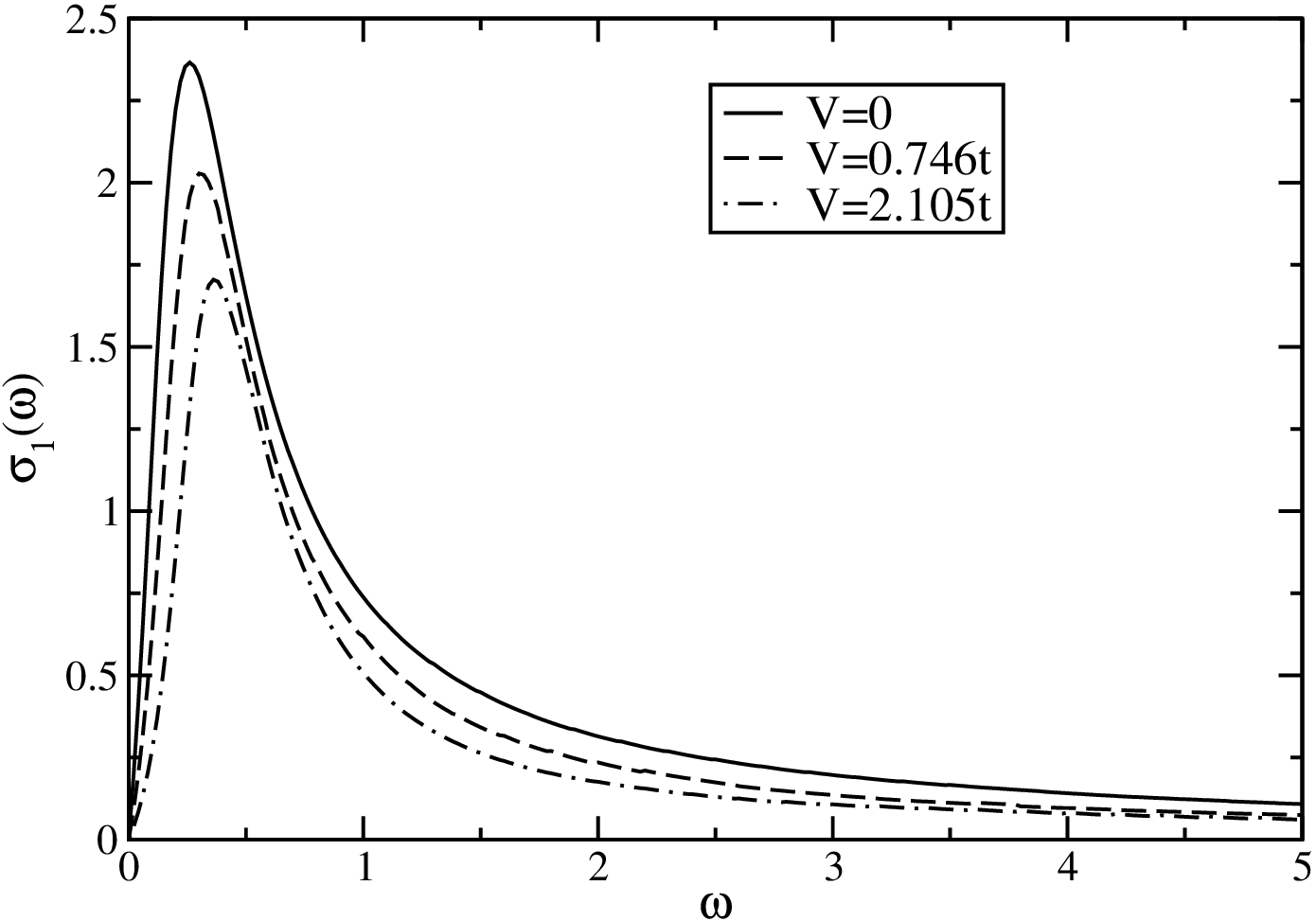}}
\caption{Optical conductivity $\sigma_1(\omega)$ for 
$\Delta=0.105t$, $U=5.263t$, $\eta = 0.2t$ ($L=64$), 
and various nearest-neighbor interactions $V$.  
}
\label{fig10}       
\end{figure}

\subsection{Nearest-Neighbor Interaction}
\label{sec:interaction}

Neglecting the long-range part of the Coulomb interaction
is difficult to justify in an insulator.
In this section we consider the effects of a nearest-neighbor
repulsion $V > 0$ in the Hamiltonian~(\ref{hamiltonian}).
(This term mimics the long-range part of the Coulomb 
interaction.)
For large enough $V$ the nature of the ground state 
of~(\ref{hamiltonian}) changes from a Mott insulator to
a CDW insulator~\cite{shi01}. 
Here we discuss only the optical conductivity
in the Mott insulating phase.

A previous DMRG investigation~\cite{nis00} of the 
model~(\ref{hamiltonian}) 
has shown that the charge gap $E_c$ increases with $V$ in the
Mott insulating phase. Our calculations confirm this result.
For $V=0$ (and $\Delta < 2t$)
the optical gap $\omega_1$ is equal to the Mott 
gap $E_c$ in the thermodynamic limit
(see Fig.~\ref{fig6} for an example).
This gap marks the onset of an excitation continuum
of unbound pairs of elementary charged excitations
(for instance, holon-antiholon pairs in the weak-coupling 
picture~\cite{con01})
which are responsible for the lowest absorption band in the
optical conductivity.  
We have found that the optical gap remains equal to the Mott gap
for non-zero but small nearest-neighbor repulsion $V$. 
Thus, the low-energy spectrum still corresponds to unbound pairs
of charged excitations.  

\begin{figure}
\resizebox{0.4\textwidth}{!}{\includegraphics{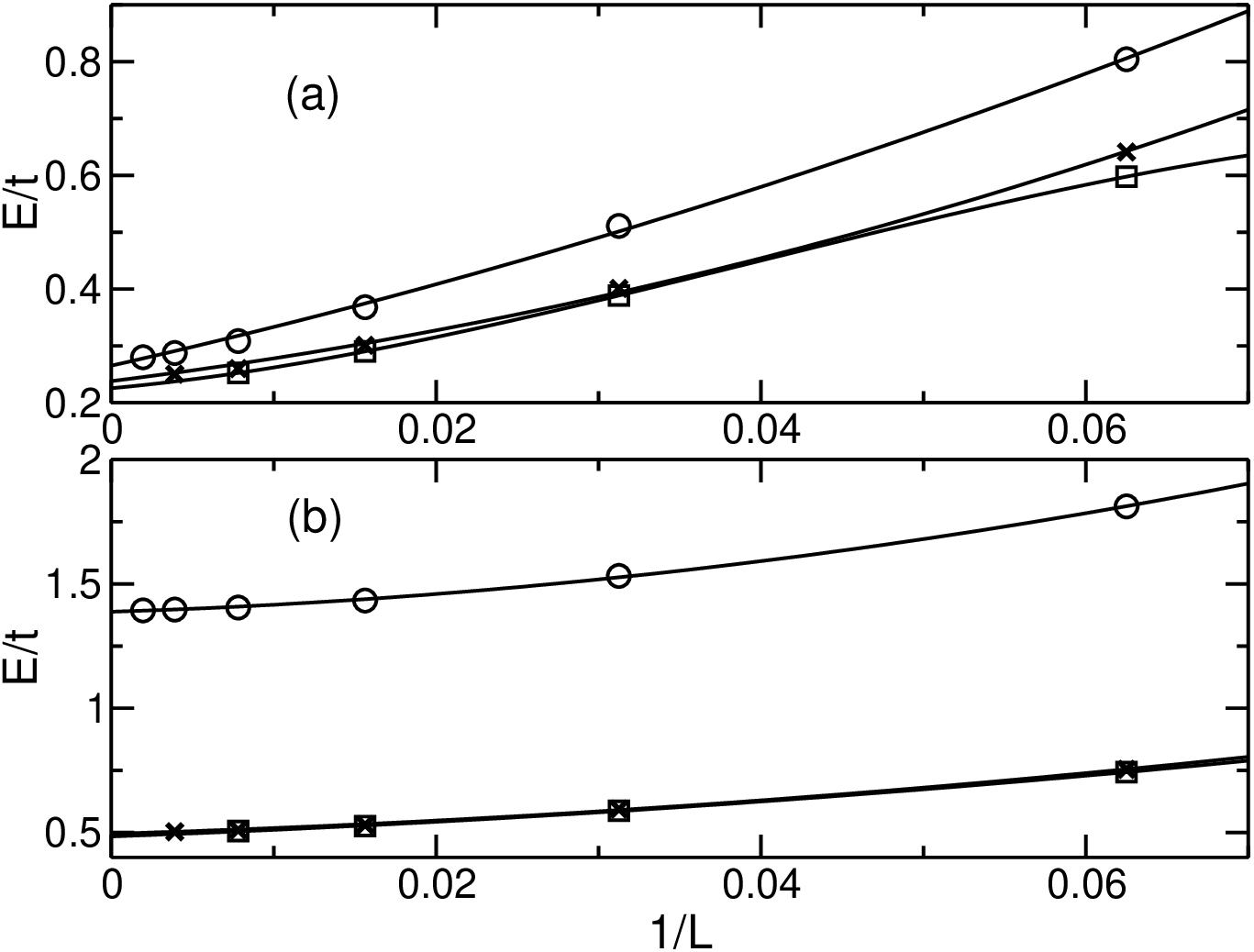}}
\caption{Mott gap $E_c$ (circle), optical gap $\omega_1$ (square), and 
position $\omega_{\rm max}$ of the spectrum maximum (cross) as a 
function of the inverse system size for
(a) $\Delta=0.105t, U=5.263t,$
and $V=2.105t$, and 
(b) $\Delta=0.353t, U=8.235t, V=3.294t$.
Lines are quadratic fits.
}
\label{fig11}       
\end{figure}

Figure~\ref{fig10} illustrates two main effects
of the nearest-neighbor repulsion on the optical spectrum
for small $V$.
First, the low-frequency peak above $E_c$ is shifted to 
higher frequency as $V$ increases because $E_c$ also increases.
Secondly, the total spectral weight decreases with increasing $V$
as described in Ref.~\cite{mil95}.

For stronger coupling $V$, however, we have found that the
optical gap extrapolates to a smaller value than the Mott gap
in the limit of an infinite chain. This can be seen 
in Fig.~\ref{fig11}. In the first example (for $\Delta=0.105t,
U=5.263t$, and $V=2.105t$) the difference $E_b = \omega_1 - E_c$
is small (about $0.04t$) but in the second example
(for $\Delta=0.353t, U=8.235t$, and $V=3.294t$) 
it is large ($E_b \approx 0.9t$). 
Moreover, the maximum of the optical spectrum is located
at a frequency $\omega_{\rm max}$
which approaches the same value as $\omega_1$
in the thermodynamic limit.
This is the signature of a excitonic peak ($\delta$-peak)
at $\omega=\omega_1$ in the optical spectrum of the Mott insulator.
(See Refs.~\cite{jec02,ess01,jec03} for a description of the
analysis that allows us to identify an 
excitonic peak in a DDMRG spectrum.)
The presence of an excitonic peak signals a fundamental change
in the nature of the lowest optical excitation.
It is now a neutral bound pair (for instance, a bound
holon-antiholon pair in field theory)
called a Mott-Hubbard exciton~\cite{ess01}.
The properties of Mott-Hubbard excitons have been
investigated using DDMRG and analytical methods
in a previous work~\cite{ess01}.
The exciton energy $\omega_1$ is smaller than the
excitation energy of the lowest unbound pair of charged
excitations in the continuum (which is $E_c$).
The difference $E_b = \omega_1 - E_c$ is the binding energy of 
the exciton. Therefore,
the first example in Fig.~\ref{fig11} corresponds
to a weakly bound (large) Mott-Hubbard exciton while the second 
example corresponds to a tightly bound (small) one.

\begin{figure}
\resizebox{0.4\textwidth}{!}{\includegraphics{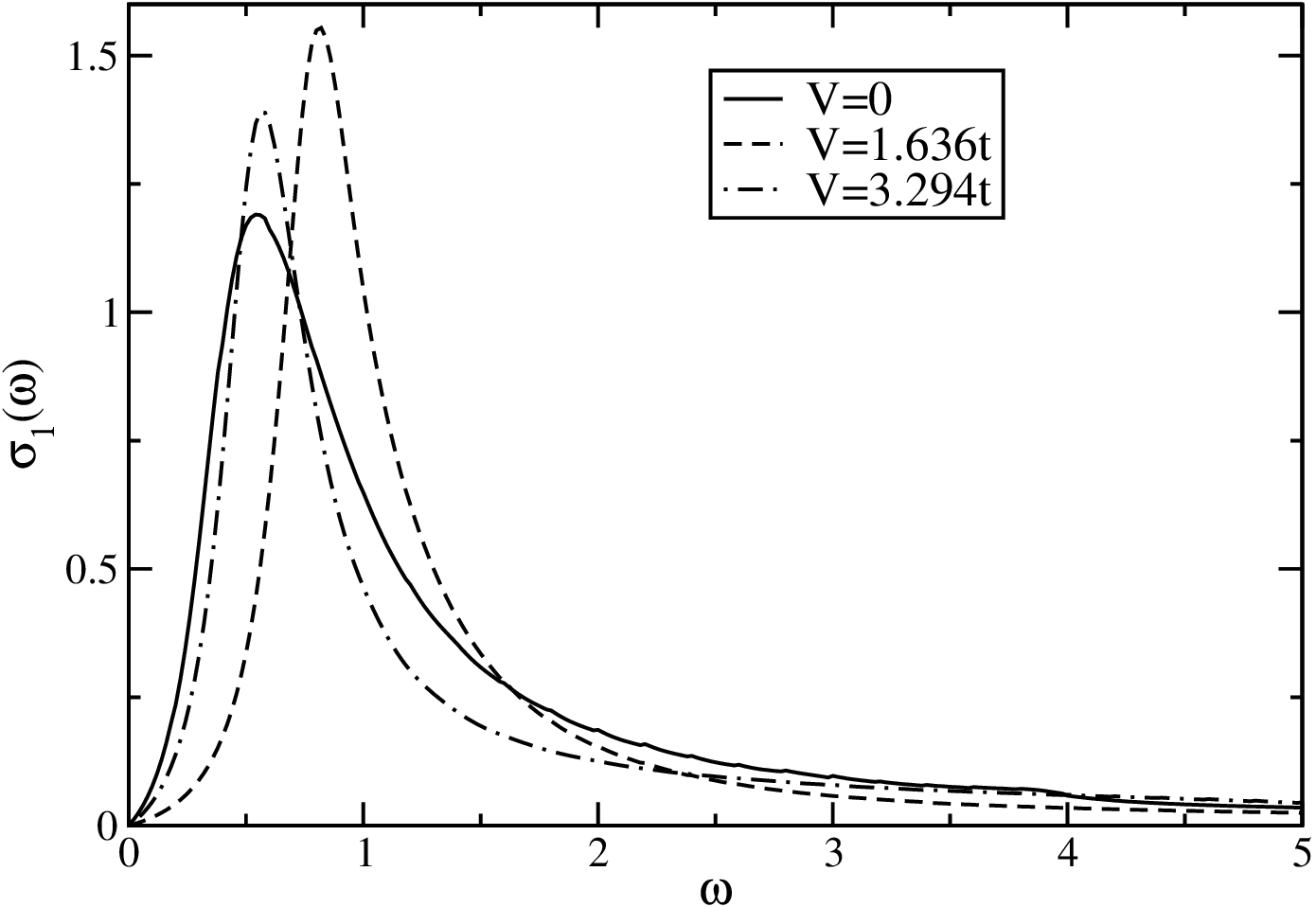}}
\caption{Optical conductivity $\sigma_1(\omega)$ for 
$\Delta=0.353t$, $U=8.235t$, $\eta = 0.2t$ ($L=64$),
and various nearest-neighbor interactions $V$.  
}
\label{fig12}       
\end{figure}

The exciton generates a $\delta$-peak at $\omega=\omega_1 < E_c$
in the optical conductivity spectrum.
As $V$ increases the exciton becomes more tightly bound (smaller)
and the optical weight is progressively transfered from
the continuum above $E_c$ to the excitonic peak.
In Fig.~\ref{fig12} one sees that the optical weight first
moves to higher 
frequency as one increases the coupling $V$ form 0 to $1.636t$
because the Mott gap $E_c$ (i.e., the continuum onset) increases.
If the coupling $V$ is increased further to $3.294t$, one finds that 
the spectral weight shifts to lower frequency because it is
transfered to the excitonic peak at $\omega_1 \approx 0.5t$
lying below the continuum onset at $E_c = 1.4t$ (which still
increases with $V$).
(Note that the gap between the exciton peak  and the continuum
is not visible in Fig.~\ref{fig12} for $V=3.294t$ because of the
broadening of the spectrum.)
The (weak) structure visible at $\omega \approx 4t$ in
the spectra calculated for $V=0$
rapidly looses  weight as $V$ increases.
In summary, we have found that
the nearest-neighbor repulsion $V$ has a significant impact
on the shape of the optical spectrum contrary to the on-site repulsion, 
but only when it is large enough to generate an exciton.

\section{Discussion}
\label{sec:discussion}

In this section we discuss the implications of our
results for the theory of the Bechgaard salts.
First, we examine which values of the model parameters 
could be appropriate for (TM)$_2$X salts.
Realistic estimates for the hopping integrals $t_1$ and $t_2$
were proposed more than ten years ago on the basis of experimental
results~\cite{ped94} and quantum-chemistry calculations~\cite{duc86}.
Using these estimates Mila~\cite{mil95} determined the model parameters
$U$ and $V$ from the reduction of the
infrared oscillator strength observed experimentally
in the Bechgaard salts. He found that relatively large
nearest-neighbor repulsions $V$ were necessary to explain
the reduction of the electron kinetic energy 
due to correlation effects.
According to Mila's analysis
appropriate model parameters are
$t_2/t_1=0.9$, $U=5t_1$, and $V=2t_1$, which correspond
to $\Delta=0.105t$, $U=5.263t$, and $V=2.105t$,
for (TMTSF)$_2$ClO$_4$
and 
$t_2/t_1=0.7$, $U=7t_1$, and $V=2.8t_1$, which correspond
to $\Delta=0.353t$, $U=8.235t$, and $V=3.294t$,
for (TMTTF)$_2$PF$_6$.
The optical spectrum obtained with DDMRG for these parameters
are shown in Figs.~\ref{fig10} and~\ref{fig12}, respectively.
As explained in Sec.~\ref{sec:interaction} we have found
that the lowest 
optical excitation is an exciton with an energy $\omega_1$
smaller than the Mott gap $E_c$ for these parameters
(see Fig.~\ref{fig11}).

More recently, Nishimoto \textit{et al.}~\cite{nis00} fitted the
Mott gap of the model~(\ref{hamiltonian}) to the
experimental optical gap to determine the model parameters.
Using the same ratios $t_2/t_1$ and $U/t_1$ as Mila 
they found that the nearest-neighbor repulsions $V$ necessary to
reproduce the optical gap data were significantly smaller than
the values given by the reduction of the oscillator strength.
According to their analysis $V=0.764t$ for (TMTSF)$_2$ClO$_4$
and $V=1.636t$ for (TMTTF)$_2$PF$_6$.
The optical spectrum obtained with DDMRG for these parameters
are also shown in Figs.~\ref{fig10} and~\ref{fig12}, respectively.
In that case we have found that
there is no exciton and an absorption continuum
due to unbound pairs of charged excitations starts at 
$\omega = \omega_1 = E_c$.

The discrepancy between these studies can be understood.
In the (TMTSF)$_2$ClO$_4$ case the kinetic energy is
a rather flat function of $V$ for the relevant parameters
$t_1, t_2, U$~\cite{mil95} while the Mott gap increases
rapidly with $V$~\cite{nis00}. Thus, the uncertainty on Mila's
value for $V$ is quite large and the value $V=0.764t$ reported
by Nishimoto \textit{et al.} is also compatible with the
experimental reduction of the oscillator strength.
Therefore, we conclude that for the salt (TMTSF)$_2$ClO$_4$ the
nearest-neighbor interaction should be close to (though somewhat
larger than) the value given by Nishimoto \textit{et al.} 
and excitons do not play any role 
in the optical excitation spectrum.  
The appropriate model
parameters for (TMTSF)$_2$ClO$_4$ are summarized in
Table~\ref{tab:1}.

In the (TMTTF)$_2$PF$_6$ case, however, the kinetic energy 
is a rather steep function of $V$ for the relevant parameters
$t_1, t_2, U$~\cite{mil95} and the value $V=1.636t$ found
by Nishimoto \textit{et al.} is only compatible with the oscillator
strength reduction for unrealistically large $U$.
As Nishimoto \textit{et al.} assumed that the experimental optical gap
corresponded to the theoretical Mott gap $E_c$, they effectively
neglected excitonic contributions to the optical spectrum.
In particular, their analysis does not take into account that the
theoretical optical gap $\omega_1$ is significantly smaller
than the Mott gap $E_c$ for large $V$ when excitons occur, as seen in 
Fig.~\ref{fig11}(b). As a result their analysis
underestimates the value of $V$.
Therefore, we conclude that for (TMTTF)$_2$PF$_6$ the
nearest-neighbor interaction should be close to (though
somewhat smaller than) the value $V=3.294t$ found by Mila
and excitons dominate the optical excitation spectrum
[at least in the framework of the model~(\ref{hamiltonian})].  
The appropriate model parameters for (TMTTF)$_2$PF$_6$ are summarized 
in Table~\ref{tab:1}.

We now examine how the main features of the optical spectrum
in (TM)$_2$X salts can be explained by the dimerized
extended Hubbard model with the parameters determined above.
Parallel to the stacks of organic molecules the optical
conductivity of (TMTSF)$_2$X salts has two distinct
components: a narrow peak at zero frequency (Drude peak)
with a very small fraction (about 1 \%) of the spectral weight
and an absorption band with most of the spectral weight
at finite energy~\cite{dre96,ves98,sch98,ves00}.
This second feature lies in the mid infrared range
above the crossover energy above
which excitations are effectively confined to a single
stack  and thus can be described by one-dimensional models.
(Obviously, the zero-energy feature always lies below
such a crossover energy and can only be described in the
framework of a three-dimensional model.)
The finite-energy feature is usually interpreted in terms
of a Mott insulator.
When  rescaled by the intensity and frequency of the spectrum 
maximum, the optical conductivity of various
(TMTSF)$_2$X salts exhibit a remarkably similar behavior~\cite{sch98}. 
In particular, a power law in the frequency dependence
\begin{equation}
\frac{\sigma_1(\omega)}{\sigma_1(\omega_{\rm max})}  = C 
\left ( \frac{\omega}{\omega_{\rm max}} \right )^{-1.3}
\label{universal}
\end{equation}
is observed over a decade in frequency $2 \omega_{\rm peak}
< \omega < 20 \omega_{\rm peak}$.  

As discussed in Sec.~\ref{sec:weak} our numerical
approach is not sufficiently accurate to confirm 
the existence of such an asymptotic power-law behavior in the 
optical spectrum of the model~(\ref{hamiltonian}).
Nevertheless, our analysis clearly indicates that if there is
a power-law behavior of $\sigma_1(\omega)$ for some frequency
range in the model~(\ref{hamiltonian}), the optical weight associated 
with
that feature must be extremely small compared to the optical weight
of the peak feature. In the experimental spectrum, however, there
is substantial optical weight in the
region where the power-law behavior is visible.
Therefore, we conclude that the universal feature~(\ref{universal})
of the optical spectrum in (TMTSF)$_2$X salts cannot be explained 
within the model~(\ref{hamiltonian}).

The well-defined mid infrared structure observed in
the optical spectrum of the Bechgaard salt (TMTSF)$_2$PF$_6$
is difficult to understand in view of the fact that 
its DC conductivity remains metallic down to very low temperature.
It seems that optical excitations are visible only for
energies much larger than the energy scale above which the system 
can be seen as metallic  (i.e., the Mott gap for
charge excitations). 
Therefore, Favand and Mila~\cite{fav96} have proposed that 
the optical gap $\omega_1$ observed in the absorption spectrum is
much larger than the Mott gap $E_c$
because of optical selection rules.
Using exact diagonalizations of small systems they 
have argued that such an effect occurs in the quarter-filled
dimerized Hubbard model~(\ref{hamiltonian}) without the
nearest-neighbor repulsion $V$.
Our analysis of this model shows that the Mott gap is smaller
than the optical gap only in the dimer case $\Delta=2t$ (see
the discussion in Sec.~\ref{sec:dimer}). For all other 
parameters $(t_1,t_2,U)$, however,
we have found that $\omega_1 = E_c$ 
($\omega_1 < E_c$  is also possible for $V > 0$ as
shown in Sec.~\ref{sec:interaction}). 
It can happen that the optical weight at the Mott gap is very small
while a very strong structure is visible in the spectrum at a
higher energy.
For instance, this occurs
in the large-dimerization limit as seen in Fig.~\ref{fig1}.
In a real material with such an absorption spectrum the weak 
low-energy band could easily be overlooked,
leading to an apparent ``optical gap'' larger than the
real gap for charge excitations $E_c$. 
For realistic parameters, however, we always find that the optical
weight is very large close to the Mott gap.
In particular, for the parameters $t_2=0.9t_1$ and $U=5t_1$
($\Delta = 0.105t$ and $U=5.263t$)
used by Favand and Mila for (TMTSF)$_2$PF$_6$
(see Table~\ref{tab:1}) we have
found that the Mott gap, the optical gap, and the maximum
of the spectrum converge to very close values in the
thermodynamic limit (see Fig.~\ref{fig6}).    
We conclude that the model~(\ref{hamiltonian}) cannot explain
the apparent discrepancy between energy scales in the
optical spectrum and the conductivity measurements for the
Bechgaard salt (TMTSF)$_2$PF$_6$.

Parallel to the stacks of organic molecules the
optical properties of (TMTTF)$_2$X salts are clearly
those of semiconductors~\cite{ves98,ves00}.
The optical spectrum displays
several strong absorption features which are attributed to
the coupling of electronic excitations with lattice vibrations.
These transitions have a higher energy than
the crossover energy above which excitations are effectively
confined to a single stack and thus can be described by a
one-dimensional model.
A remarkable property of the (TMTTF)$_2$X salts is that
the optical gap is smaller than the Mott gap
determined by photoemission experiments. For (TMTTF)$_2$PF$_6$
one observes
that the strongest structure in the optical spectrum is
a relatively sharp peak at an energy (about 100 meV) significantly
smaller than the Mott gap (about 200 meV).
Obviously, this can be interpreted as the signature of an excitonic
transition below the gap for charged excitations.
Thus, experimental observations are compatible with
the theoretical prediction of excitons in (TMTTF)$_2$PF$_6$
based on the model~(\ref{hamiltonian}) and  Mila's
estimation~\cite{mil95} for the appropriate parameters (especially,
the strength of the nearest-neighbor interaction $V$).
Quantitatively, we find $E_c\approx 160$ meV and an exciton
energy of $\omega_1 \approx $ 60 meV using
the parameters in Table~\ref{tab:1}. 
These energies are in very
satisfactory agreement with the experimental values.
Therefore, we conclude that excitons are present in the
optical spectrum of (TMTTF)$_2$PF$_6$ [and probably other
(TMTTF)$_2$X salts] and explain the observation of
absorption features below the gap
measured with photoemission spectroscopy.
It would be interesting to have a direct experimental evidence
for the presence of excitons in those salts. For instance,
one could investigate the electro-absorption spectrum to
demonstrate the presence of excitons
as it was done
for another quasi-one-dimensional material,
polydiacetylene~\cite{weiser}.


In conclusion, we
have investigated the optical conductivity of the
one-dimensional dimerized extended Hubbard model at quarter filling
using the dynamical density-matrix renormalization group.
We have found that the dimerization amplitude and the nearest-neighbor
repulsion (if strong enough to form excitons) determine the main 
features of the optical spectrum. Besides its influence on the
energy scale set by the Mott gap, the on-site repulsion plays
a minor role only.
Our study shows that this model cannot explain the optical
spectrum in the Bechgaard salts (TMTSF)$_2$X.
It also shows that excitons probably contribute
to the optical spectrum in the (TMTTF)$_2$X salts.

\subsubsection*{Acknowledgments}

We are grateful to S. Nishimoto,
F. Gebhard, and F. Mila
for helpful discussions. H.B. acknowledges
support by the Optodynamics Center of the
Philipps-Universit\"{a}t Marburg.

\end{document}